# CsCl seed layer homogenizes co-evaporated perovskite growth for high-efficiency fully textured perovskite-silicon tandem solar cells

Viktor Škorjanc[*,1] (0000-0002-2438-0090), Stefanie Severin[1], Alexander Veber[1,2] (0000-0003-4894-5639), Mauricio J. Prieto[3] (0000-0002-5087-4545), Liviu C. Tănase[3] (0000-0002-4177-5676), Aleksandra Miaskiewicz[1], Sebastian Weitz[1], Jing-Wen Hsueh[3] (0009-0009-3482-5776), Mohamad-Assaad Mawass[3] (0000-0002-6470-2920), Lucas de Souza Caldas[3] (0000-0002-5499-4712), Suresh Manyarasu[1], Philippe Holzhey[1] (0000-0003-3688-1607), Erik Wutke[1], Stepan Demchyshyn[1], Matthew R. Leyden[1] (0000-0002-7485-3351), Angelika Harter[1], Roberto Felix Duarte[1], Jona Kurpiers[1], Philipp Wagner[1], Bernd Stannovski[1], Ljiljana Puskar[1], Roland Mainz[1], Daniel Abou-Ras[1], Thomas Schmidt[3] (0000-0003-4389-2080), Lars Korte[1] (0000-0002-9207-9048), Marcel Roß[*,1], Steve Albrecht[*,1,4] (0000-0001-9962-9535)

1. Helmholtz-Zentrum Berlin für Materialien und Energie, Hahn-Meitner-Platz 1, 14109 Berlin, Germany

2. Department of Chemistry, Humboldt-Universität zu Berlin, Brook-Taylor-Str. 2, 12489 Berlin, Germany

3. Department of Interface Science, Fritz Haber Institute of the Max Planck Society, Faradayweg 4-6, 14195 Berlin, Germany

4. Faculty of Electrical Engineering and Computer Science, Technical University Berlin, Marchstraße 23, 10587 Berlin, Germany

*Corresponding Authors: viktor.skorjanc@helmholtz-berlin.de, marcel.ross@helmholtz-berlin.de, steve.albrecht@helmholtz-berlin.de

# Abstract

Monolithic perovskite-silicon tandem solar cells experienced a significant increase in efficiency, making them viable for industrial applications. Among the various scalable and industry-compatible metal halide perovskite deposition techniques, co-evaporation stands out as particularly well-suited for perovskite-silicon tandem solar cells due to its ability to conformally cover textured silicon bottom cells. Solution-processed [2-(3,6-Dimethoxy-9*H*-carbazol-9-yl)ethyl]phosphonic acid (MeO-2PACz) is commonly used as a hole-transporting material for co-evaporated metal halide perovskites. However, we show that it covers the textured surface of silicon bottom cells unevenly, impacting the film growth and leading to the formation of residual $PbI_2$ at the buried interface. The present study reveals via X-ray photoemission electron microscopy (XPEEM) and infrared scattering-type scanning near-field optical microscope (IR s-SNOM) that a CsCl seed layer fosters organic precursor incorporation across the MeO-2PACz/perovskite interface, even in the areas with a thin MeO-2PACz layer, thereby preventing the formation of interfacial $PbI_2$ and leading to larger apparent grains. The improvement of the metal halide perovskite film quality on the MeO-2PACz/perovskite interface and the bulk perovskite film led to 30.3% (29.7% certified) efficient perovskite-silicon tandem solar cell. The present work highlights the importance of a seed layer for a robust growth of co-evaporated metal halide perovskite and represents an important milestone for the transfer of perovskite-silicon tandem solar cells from laboratory to industry.

# 1. Introduction

Monolithic perovskite–silicon tandem solar cells (PSTs) have achieved a rapid increase in power conversion efficiency, reaching 34.85%, a substantial improvement over single-junction devices.[1] This remarkable performance underscores the growing importance of advancing PST industrialization. Metal halide perovskite (MHP) vacuum deposition is a solvent-free up-scalable deposition method which is particularly applicable for PSTs as it allows for conformal coverage of textured silicon bottom cells.[2] Vacuum deposition also offers excellent thickness control and enables the formation of vertical compositional gradients within the MHP film.[3] In addition, vacuum-deposited MHP films are free of residual solvents that can compromise device stability.[4–7] However, its higher equipment costs and lower throughput have limited its adoption, resulting in significantly less research compared to solution-based techniques. Consequently, the highest reported power conversion efficiency (PCE) for fully evaporated PSTs is 27.4%.[8] Figure S 1 shows summary of published PST efficiencies as a function of publication date, categorized by deposition method and bottom-cell texture.

An important milestone in the development of PSTs was the introduction of carbazole-based self-assembled layers (often referred to as self-assembled *monolayers*, SAMs, although in practice they form mostly layers exceeding a single monolayer) as hole transport layer (HTL).[9] Owing to their simple deposition, effective defect passivation, and tunable energy-level alignment, carbazole-based SAMs were rapidly adopted by the PST community.[10–15] However, carbazole-based SAMs, with a hydrophilic phosphonic acid head and hydrophobic carbazole tail, are prone to micelle formation,[16] which can cause uneven substrate coverage. Mitigation strategies include co-solvents,[16] co-adsorbents,[17] and post-deposition solvent treatments.[18,19] Their coverage and thickness also depends on the underlying transparent conductive oxide (TCO) substrate, with smoother amorphous TCOs typically favoring more homogeneous films.[20,21] In addition, residual unbound SAM molecules can remain after deposition, particularly relevant for vacuum-processed MHPs where they are not removed by the MHP precursor solution. These species modify interfacial growth and composition, highlighting the importance of controlling SAM coverage and thickness.[22–24]

Achieving uniform SAM coverage on textured silicon substrates remains a particular challenge. Park et al. demonstrated a 25.3% efficient single-junction PSC on glass/fluorine-doped tin oxide (FTO) substrate with pyramid texture up to 200 nm high by improving the adsorption of phosphonic acids through a 3-mercaptopropionic acid (3-MPA) co-adsorbent combined with a [2-(9*H*-carbazol-9-yl)ethyl]phosphonic acid (2PACz) SAM, which was spin-coated and subsequently washed with ethanol.[17] Jia et al. achieved a 34.6% efficient PST using an asymmetric spin-coated SAM molecule with reduced steric hindrance, enabling improved coverage on 0.5–1 μm pyramid textures.[25] However, both approaches relied on solution-processed MHPs and smaller pyramid textures, which are not compatible with standard photovoltaic manufacturing process. Kore et al. reported 28.6% efficient PSTs using evaporated 2PACz SAMs combined with hybrid-deposited MHPs on larger 1–2 μm pyramid textures.[26] Zhang et al. demonstrated 33% efficient PSTs with spin-coated 2PACz and MHP on industrially relevant textured silicon with pyramid size exceeding 2 μm by introducing silica nanospheres that cover the pyramid valleys, enabling more uniform coverage.[27] Demonstration of >30% efficient PSTs deposited on micron-sized pyramids using up-scalable perovskite deposition technique that does not include spin-coating is still lacking.

In this study, we use advanced synchrotron-based microscopic analysis techniques to show that inhomogeneities in the [2-(3,6-dimethoxy-9*H*-carbazol-9-yl)ethyl]phosphonic acid (MeO-2PACz) layer play an important role in driving variations in MHP growth. First, we apply infrared scattering-type scanning near-field optical microscopy (IR s-SNOM) to probe the bonding of self-assembled layers on the TCO substrate with nanoscale resolution, revealing clear differences between a true monolayer and thicker molecular layers. We then investigate how the inhomogeneities in MeO-2PACz coverage affect MHP growth on a planar substrate. We use x-ray photoelectron emission microscopy (XPEEM) on ultra-thin MHP films (5–20 nm) to probe local inhomogeneities in film formation, and on delaminated thicker films to access buried interfaces and evaluate interfacial inhomogeneities post-deposition.

Building on the insights for the planar substrate, we extend our analysis to industrially relevant fully textured silicon bottom cells. Here, we show that residual MeO-2PACz solution tends to accumulate in the valleys of the random pyramids, leading to comparably thinner coverage on the pyramid facets. This leads to poor incorporation of the organic precursor at the pyramid facets which are therefore $PbI_2$-rich at the buried interface. A CsCl seed layer can mitigate these inhomogeneities by promoting a more homogeneous lateral incorporation of the organic component. As a result, the CsCl seed layer effectively prevents the formation of interfacial $PbI_2$ and leads to improved film uniformity, even on challenging textured substrates, leading to an improvement in fill factor (FF) and open-circuit voltage ($V_{OC}$) and reduced variability in band gap, enabling PSTs with PCE of 30.3% (29.7% certified), the highest value for PSTs with fully vacuum-processed MHP layer.

# 2. Results and discussion

## 2.1 Inhomogeneous MeO-2PACz coverage

To study co-evaporated film formation on silicon bottom cells, we performed compositional analysis on MHP films of various thicknesses deposited on polished silicon substrates coated with a metal-doped $In_2O_3$-based TCO layer (abbreviated as IMO:H), which mimic a planar silicon bottom cell. Planar substrates were chosen instead of textured ones as compositional analysis on textured surfaces is significantly more challenging. In the case of XPEEM measurements used in this study, sharp features such as the pyramid tips of a textured wafer can distort the local electric field, with a negative impact on the resolution of the measurement. Prior to MHP co-evaporation, MeO-2PACz HTL was spin-coated on top of the substrate, however, the low roughness led to partial coalescence of the MeO-2PACz film during annealing.

To determine the lateral distribution of molecular bond vibrations in MeO-2PACz film, we applied infrared scattering-type scanning near-field optical microscopy (IR s-SNOM) for nanoscale IR spectroscopy. We first obtained a topography and optical amplitude images, which are recorded simultaneously. The topography image (Figure 1a) reveals three distinct regions, which we assign to thin MeO-2PACz (height < 8 nm), thick MeO-2PACz (10 nm < height < 15 nm) and MeO-2PACz agglomerate (height > 15 nm). The height range in Figure 1a was adjusted to highlight the difference in height between the thick and thin MeO-2PACz layer. Figure S1a shows the entire height range. Figure 1b shows the height profile across the boundary between thick and thin MeO-2PACz regions, taken along the line marked in the topography image (Figure 1a). The step height is approximately 6 nm, which corresponds to several molecular layers (see Supplementary Note 1 for details). The optical amplitude image (Figure S 2b), which was obtained in the white-light mode using broadband IR synchrotron radiation is anti-correlated to the topography image, indicating that the higher IR absorption is caused by thicker MeO-2PACz.

To confirm that the three regions indeed correspond to thin, thick and agglomerated MeO-2PACz, we acquired local IR spectra from areas of about ~25 × 25 $nm^2$ (determined by the radius of the AFM tip) at three distinct points marked with crosses on the topography image (Figure 1a). All three positions show multiple bands corresponding to MeO-2PACz in the nano-IR spectra (Figure 1c), such as 1579 $cm^{-1}$ (carbazole modes) and 1476 $cm^{-1}$ (carbazole, $CH_3$ and $CH_2$ modes).[12,28] A detailed assignment of MeO-2PACz vibrational modes can be found in ref. 28. The most intense signal is observed at the point with agglomerated MeO-2PACz, consistent with increased material thickness as seen in the topography and optical amplitude images. The spectra from the agglomerate and thick MeO-2PACz regions closely match IR spectra reported in the literature measured via common far-field techniques for bulk MeO-2PACz powder.[12,28–30] Spectral differences between the thick and thin MeO-2PACz layers indicate that the thin MeO-2PACz layer is consistent with a self-assembled monolayer, as detailed in Supplementary Note 2. Possible influence of MeO-2PACz inhomogeneities on solar cell device performance is discussed in Supplementary Note 3.

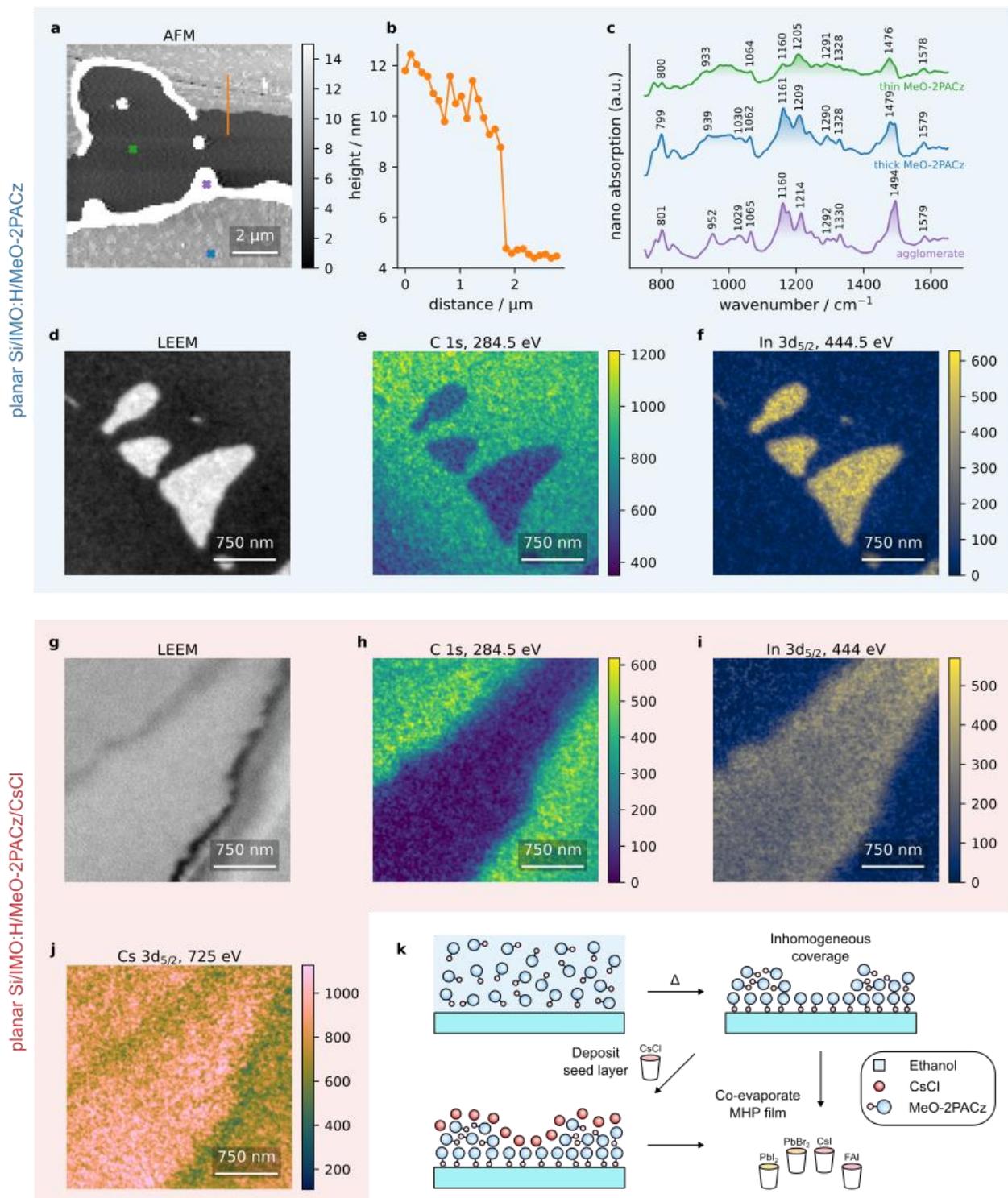

**Figure 1**: CsCl seed layer covers areas with inhomogeneous MeO-2PACz coverage. **a**, Topography image of a planar Si/IMO:H/MeO-2PACz sample with inhomogeneous MeO-2PACz coverage. Orange line marks where line profile was determined. Crosses mark spots where IR s-SNOM spectra were measured. Color scale shows height in nm. **b**, Line profile of line shown in panel a. Top of the line is shown on the left. **c**, IR s-SNOM spectra of three spots marked on panel a. **d-j**, LEEM/XPEEM analysis of inhomogeneous MeO-2PACz coverage on planar Si/IMO:H/MeO-2PACz (**d-f**) and planar Si/IMO:H/MeO-2PACz/CsCl (**g-j**) sample. **d,g**, LEEM image. **e,f,h-j**, XPEEM image of C 1s (**e,h**), In 3d$_{5/2}$ (**f,i**) and Cs 3d$_{5/2}$ (**j**) core levels. Binding energy is indicated in the title. The lower intensity of the signal in the bottom left corner of image e can be attributed to the non-uniform illumination profile of the XPEEM beam. Color scale shows XPEEM signal intensity. Images in d-f and g-j were measured on the same area of a sample. **k**, Schematic illustrating the formation of MeO-2PACz inhomogeneities and their subsequent coverage by the CsCl seed layer.

Given the strong influence of MeO-2PACz thickness on MHP film formation,[24,31] unevenly covered substrates serve as a valuable model system for studying how variations in SAM layer thickness affect MHP growth. MHP co-evaporation is particularly well suited for this purpose, as it does not dissolve residual MeO-2PACz during deposition and enables the formation of ultra-thin films to directly probe the interface, an approach not possible with solution processing. These findings are especially relevant for textured silicon bottom cells, where we observe similar inhomogeneities, even when using lower MeO-2PACz concentrations and adding *N,N*-dimethylformamide (DMF) as a co-solvent to suppress micelle formation and improve coverage uniformity,[16] which will be discussed below.

To investigate how inhomogeneous MeO-2PACz coverage influences MHP film formation, we performed spectro-microscopic analysis using low energy electron microscopy (LEEM) and XPEEM. Measurements were carried out on bare planar Si/IMO:H/MeO-2PACz substrates, as well as on samples with 5 nm and 20 nm thick MHPs, and on the buried interface of delaminated 100 nm-thick MHP films, with and without a CsCl seed layer. We refer to samples without a seed layer as reference samples. In this study, LEEM contrast, determined by the reflectivity of low-energy electrons interacting with different surface layers, is primarily used to distinguish morphological variations. XPEEM provides laterally resolved XPS spectra and spatial maps of core-level signal intensity, thus of the presence and bonding environment of atomic species. Additional details on the LEEM and XPEEM techniques are provided in the Methods section.

To complement the spectro-microscopic measurements, Figure S 3a shows a top-view scanning electron microscope (SEM) image revealing inhomogeneous SAM coverage. This differences are also visible in the morphology of a sample with 5 nm co-evaporated MHP (Figure S 3b), as well as the buried interface of a 100 nm thick delaminated MHP sample (Figure S 3c), indicating that inhomogeneities in MeO-2PACz coverage influence MHP film formation and growth. Differences are less pronounced in equivalent samples with a CsCl seed layer (Figure S 3d-f).

Inhomogeneities in Si/IMO:H substrate coverage by MeO-2PACz, observed in IR s-SNOM analysis, are also visible in LEEM image (Figure 1d). XPEEM images show that the bright region in LEEM corresponds to weak C 1s (Figure 1e) and strong In $3d_{5/2}$ (Figure 1f) core level signal. The opposite is true for the dark region. If we assume that the C 1s signal originates from MeO-2PACz, and the In $3d_{5/2}$ originates from $In_2O_3$ in IMO:H, we can assign the bright region in LEEM to thin, and the dark region to thick MeO-2PACz layer. It is important to note that the C 1s signal at ~284.5 eV can, in addition to MeO-2PACz, also originate from adventitious carbon originating from atmospheric contamination. While adventitious carbon may indeed account for a part of the C 1s signal, we would expect it to cover the substrate rather homogeneously, as opposed to the XPEEM C 1s core level image which reveals a strong and well-defined contrast in C 1s signal distribution. In addition, the substrates were transferred and loaded in $N_2$ atmosphere for XPEEM measurements to minimize the generation of adventitious carbon on the sample surface. Strong In $3d_{5/2}$ signal of $In_2O_3$ originating from the region with thin MeO-2PACz indicates that the MeO-2PACz layer thickness is comparable to or thinner than the XPEEM information depth at the used kinetic energy (~150 eV), consistent with nano-IR spectra showing that the thin MeO-2PACz layer resembles a monolayer (See Supplementary Note 4 for details). Supplementary Note 5 describes the secondary electron cutoff spectra (Figure S 4a), core-level XPS spectra obtained by averaging the signal intensity in the regions assigned to thin and thick MeO-2PACz (Figure S 4b-e) and XPEEM images of other constituent elements (Figure S 5).

In summary, we show that the C 1s and In $3d_{5/2}$ core level signals can be used to identify the inhomogeneities in MeO-2PACz coverage, with regions with thick MeO-2PACz characterized by a strong C 1s and weak In $3d_{5/2}$ core level signal.

## 2.2  CsCl seed layer distribution

Motivated by our previous study,[32] we investigated how a CsCl seed layer covers areas with inhomogeneous MeO-2PACz thickness by analyzing the morphology and elemental distribution of Si/IMO:H/MeO-2PACz samples coated with a 10 nm CsCl layer. Figure 1g shows a LEEM image of a region with inhomogeneous MeO-2PACz coverage. A diagonal stripe, running from the bottom left to the top right, is visible. We assign the feature in the bottom right edge of the stripe, resembling an elevated domain, to a MeO-2PACz agglomerate. In XPEEM, the diagonal stripe exhibits weak C 1s (Figure 1h) and strong In $3d_{5/2}$ (Figure 1i) signals, consistent with a region with thin MeO-2PACz. The Cs $3d_{5/2}$ signal (Figure 1j) shows that cesium covers both thick and thin MeO-2PACz areas relatively uniformly. However, slightly lower Cs signal intensity is observed at the MeO-2PACz agglomerate in the bottom right corner and at the edges of the thin MeO-2PACz region. The SEM image of a CsCl-coated sample (Figure S 6) reveals that the seed layer forms interconnected islands ~35–50 nm in diameter, too small to distinguish in XPEEM at the field of view used. C 1s and In $3d_{5/2}$ signals therefore remain visible despite the nominal 10 nm seed layer thickness which is considerably higher than the estimated information depth of 2.2 nm (see Methods for details), as the substrate is not fully covered. Core-level XPS spectra for the CsCl-coated sample are shown in Figure S 7 and discussed in Supplementary Note 6. In summary, we show that CsCl seed layer covers both regions with thin and thick MeO-2PACz layer but less on agglomerated regions. Formation of inhomogeneities and coverage by the CsCl seed layer are shown schematically in Figure 1k.

## 2.3 Metal halide perovskite growth

Next, to assess how uneven MeO-2PACz coverage influences MHP growth, we deposited $FA_{0.8}Cs_{0.2}PbI_{2.7}Br_{0.3}$ MHP films of various thicknesses by four-source co-evaporation onto a Si/IMO:H/MeO-2PACz reference sample and a Si/IMO:H/MeO-2PACz/CsCl sample. All films were prepared in the same deposition run to reduce variations caused by process drift.

We first analyzed a 5 nm thick MHP film on a planar Si/IMO:H/MeO-2PACz reference substrate. The LEEM image (Figure 2a) reveals a bright region distinct from the rest of the film. Figure S 8 shows XPS spectra of the constituent elements, while Figure 2b–d and Figure S 9 show XPEEM images. XPEEM images show that the bright region in LEEM exhibits a low C 1s signal (Figure 2b) and a high I $3d_{5/2}$ signal (Figure 2c). The In $3d_{5/2}$ signal (Figure 2d) is mainly discernible along the edges of the carbon-poor region and at smaller, distinct points elsewhere on the film. These points also exhibit a strong C 1s (Figure 2b) and a weak Pb $4f_{7/2}$ signal (Figure S 9b). We assign these regions to pinholes: areas where no MHP precursor film was deposited or where the film is extremely thin. The C 1s signal in these regions likely originates from interfacial MeO-2PACz. Aside from the pinholes, the In $3d_{5/2}$ signal is slightly stronger in the region with weak C 1s signal.

With a thin MHP film atop the MeO-2PACz layer, the carbon signal can originate not only from MeO-2PACz and adventitious carbon, as discussed earlier, but also from the organic precursor FAI. MeO-2PACz is the more likely carbon signal source due to the XPS peak maximum position, as detailed in Supplementary Note 7. XPS spectra and XPEEM images of other elements are discussed in Supplementary Note 8.

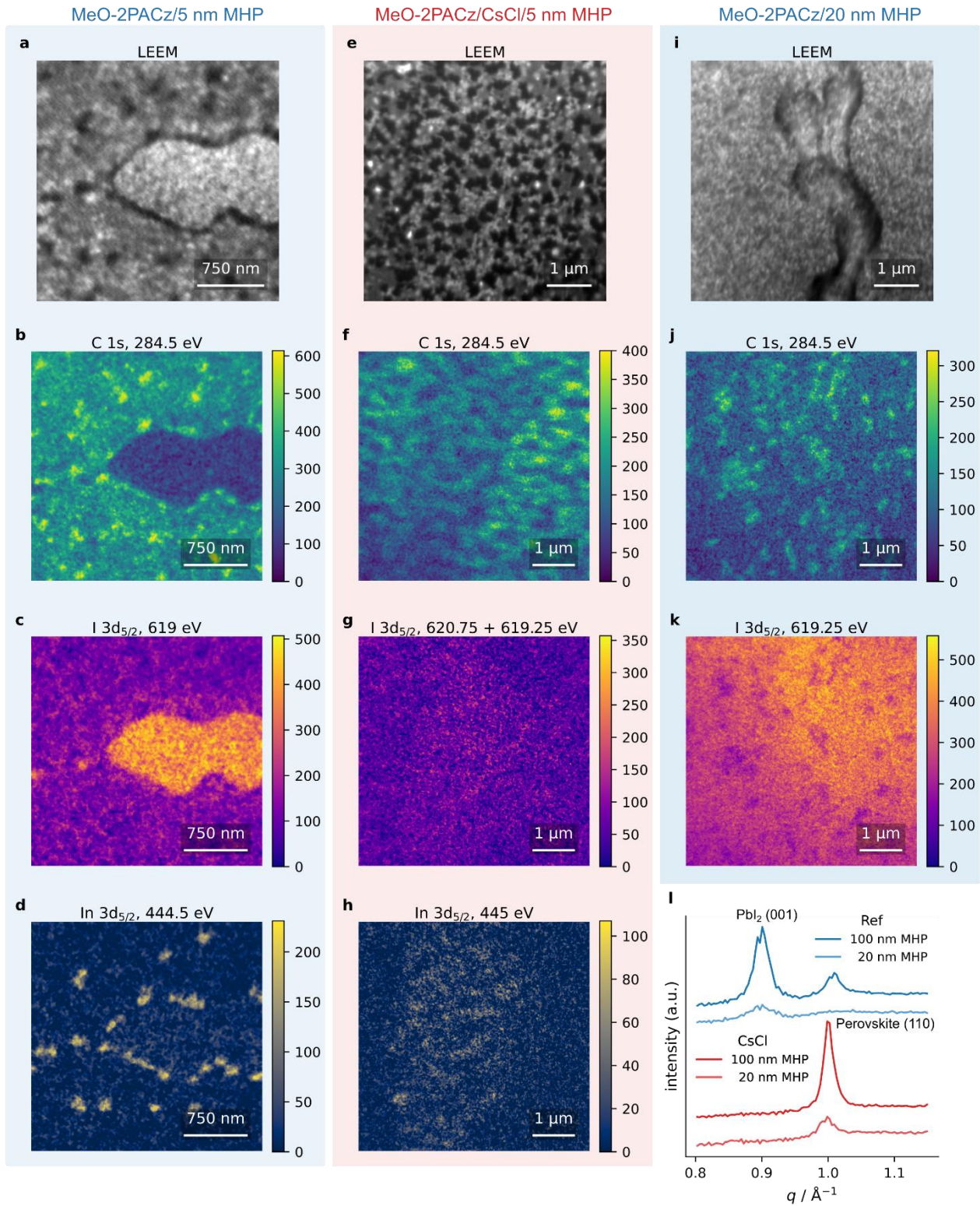

**Figure 2**: CsCl fosters organic precursor incorporation in areas with thin MeO-2PACz layer. **a-d** 5 nm MHP reference film. **e-h**, 5 nm MHP film grown on CsCl seed layer. **i-k**, 20 nm MHP reference film. **a,e,i**, LEEM image. **b-d,f-h,j,k** XPEEM image of C 1s (**b,f,j**), I $3d_{5/2}$ (**c,g,k**) and In $3d_{5/2}$ (**d,h**) core levels. Color scale shows XPEEM signal intensity. **l**, XRD patterns of 20 and 100 nm thick MHP films on reference and CsCl substrate.

Compared to the reference substrate (Figure 2a), the region with thin MeO-2PACz on the sample with 5 nm MHP grown on CsCl seed layer is less distinct in the LEEM image (Figure 2e). However, it becomes visible in the C 1s (Figure 2f) and In $3d_{5/2}$ (Figure 2h) core-level images, where it is characterized by lower carbon and higher indium intensity. When comparing the carbon distribution on the reference sample to the sample with CsCl, the most significant difference is in the thin MeO-2PACz region: in the reference sample, the C 1s signal was essentially absent (Figure 2b), whereas in the CsCl sample, C-rich islands are present. Since the CsCl sample without MHP also showed low C 1s intensity in the region with thin MeO-2PACz, it is unlikely that the observed carbon originates from MeO-2PACz or from contamination during CsCl evaporation. We therefore attribute the C 1s signal in these islands to material deposited during MHP evaporation, likely from FA$^+$ or an FAI degradation product, even though the measured binding energy is lower than expected for C=N or C≡N bonds. This supports the conclusion that CsCl promotes incorporation of the organic precursor, consistent with findings in our previous work.[32] The C 1s distribution is similar in the thick MeO-2PACz region, with C-rich islands forming, but the signal intensity is higher than in the thin region. This higher intensity could result from combined contributions of MeO-2PACz and co-evaporated MHP to the C 1s signal, or from a synergistic effect of CsCl and thick MeO-2PACz that enhances incorporation of the organic precursor. As shown in Figure S 6, CsCl does not fully cover the substrate, but rather forms islands, meaning that the underlying MeO-2PACz can still influence film growth.

XPS spectra of the constituent elements are presented in Figure S 11. For I $3d_{5/2}$, the binding energy in the thin MeO-2PACz region is 1 eV lower than in the surrounding film (Figure S 11a). Both peak positions were imaged (Figure S 12a,b), and the iodine distribution shown in Figure 2g represents the sum of the two images to represent the overall iodine distribution. Iodine distribution is much more homogeneous compared to the reference sample (Figure 2c), indicating more homogeneous film formation. Other XPS spectra and XPEEM images of other elements (Figure S 12c-j) are discussed in Supplementary Note 9.

To assess how MeO-2PACz layer thickness influences the composition of thicker MHP films, we examined elemental distributions in a reference sample with a 20 nm MHP layer. Even though the region with thin MeO-2PACz is still visible as a distinct morphological feature in the LEEM image (Figure 2i), the distribution of C 1s (Figure 2j) and I $3d_{5/2}$ (Figure 2k) does not significantly differ between the region with thin and thick MeO-2PACz. This is in stark contrast to the 5 nm thick MHP on reference sample where C 1s signal was absent in the region with thin MeO-2PACz (Figure 2b), and indicates the inclusion of the organic precursor into the film, as the signal clearly does not originate from MeO-2PACz in the region with thin MeO-2PACz. This shows that, after an initial inorganic layer is formed on top of the area with a thin MeO-2PACz layer, this inorganic layer acts as a seed layer and allows for incorporation of organic precursor into the film. We note that the core-level intensity distribution is not homogeneous, showing pronounced carbon-rich domains. Images of the core-level signal distribution of other elements are shown in Figure S 13a-c. The substrate- and MeO-2PACz-related core levels O 1s, P 2p and In $3d_{5/2}$ were not discernible for 20 nm thick MHP films.

XRD patterns of as-deposited films (Figure 2l) show that for 20 nm films, the reference sample exhibits a peak at $q = 0.9$ Å$^{-1}$, corresponding to the (001) lattice plane of PbI$_2$, whereas the CsCl sample shows a peak at $q = 1.0$ Å$^{-1}$, corresponding to the (110) lattice plane of MHP. For the 100 nm reference film, the PbI$_2$ peak at $q = 0.9$ Å$^{-1}$ remains dominant, although a weaker MHP peak at $q = 1.0$ Å$^{-1}$ is also present. In contrast, the 100 nm CsCl film displays only the MHP peak.

These results indicate that in the reference sample, MHP phase formation during co-evaporation is delayed, occurring only after a significant thickness is reached, with the initial film layer consisting mainly of PbI$_2$. Once formed, this PbI$_2$-rich interface appears to favor higher PbI$_2$ incorporation during co-evaporation, as indicated by the strong PbI$_2$ peak in the 100 nm thick reference sample. In samples with CsCl seed, MHP forms directly at the interface, and this initial MHP layer promotes continued MHP growth. The differences in early film growth therefore influence film formation even at higher film thicknesses.

## 2.4 Composition, morphology and optoelectronic properties of the buried interface

While analyzing thin layers of MHP film can give valuable insight into the initial film formation close to the substrate surface, it is still possible that, during the deposition process, the precursors diffuse,[33] altering the composition of the MHP/MeO-2PACz/IMO:H interface. To analyze the MHP composition at the buried interface after the deposition, we delaminated a 100 nm thick as-deposited MHP film by using a UV-curable glue (see Methods for more details) and measured XPEEM on the exposed surface of the delaminated film.

In the reference film, inhomogeneities in morphology are visible in the LEEM image (Figure 3a). The C 1s XPEEM image (Figure 3b) reveals a diagonal stripe with reduced intensity, corresponding to the region with thin MeO-2PACz as discussed in the following. The top-left and bottom-right corners exhibit strong C 1s signals, likely originating from residual MeO-2PACz that remained on the MHP surface after delamination of the thick MeO-2PACz region. A localized, intense C 1s signal in the top-right corner is attributed to an agglomerate of MeO-2PACz. The I $3d_{5/2}$ image (Figure 3c) is anti-correlated to the C 1s image, with higher intensity in the thin MeO-2PACz region, consistent with observations for the 5 nm MHP reference film. This suggests that diffusion during deposition does not significantly modify the quantity of the organic component present at the interface, despite evidence of organic precursor incorporation in MHP grown on the thin MeO-2PACz region for the 20 nm MHP case shown previously (see C 1s XPEEM image in Figure 2j). XPS spectra (Figure S 14) and XPEEM images of additional core levels (Figure S 15) are discussed in Supplementary Note 10.

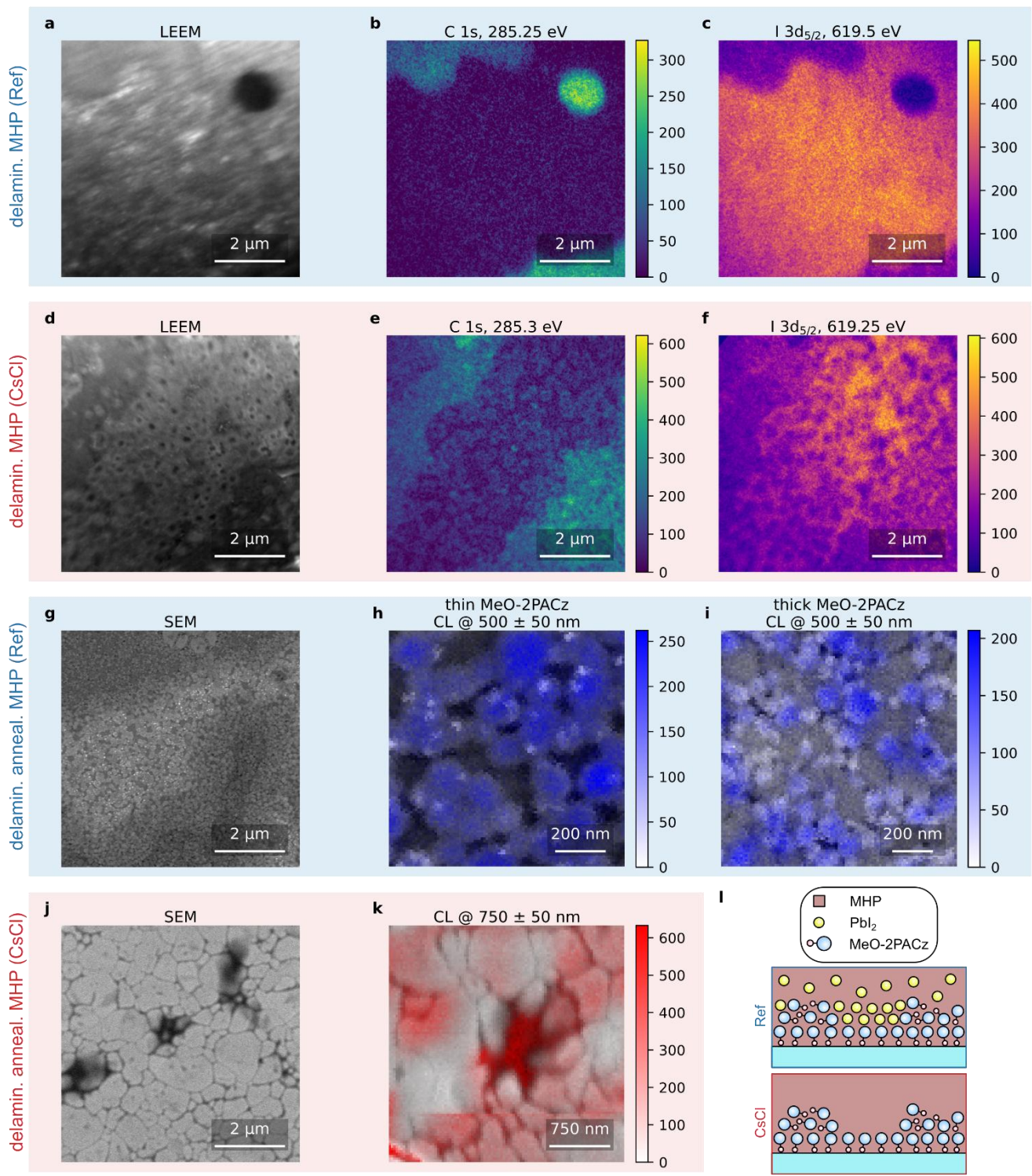

**Figure 3**: CsCl seed layer suppresses formation of interfacial PbI$_2$. **a-f**, Buried interface of 100 thick, as-deposited MHP. **a-c**, Reference sample. **d-f**, CsCl sample. **a,d**, LEEM image. **b,c,e,f**, XPEEM image of C 1s (**b,e**) and I 3d$_{5/2}$ (**c,f**) core level electrons. Color scale shows XPEEM signal intensity. **g-k**, Buried interface of a 500 nm thick MHP film annealed prior to delamination. SEM images of reference (**g-i**) and CsCl (**j,k**) sample. Images in **h,i,k**, are overlaid with cathodoluminescence (CL) images measured in the range of 500 ± 50 nm (**h,i**) and 750 ± 50 nm (**k**).

The LEEM image of the CsCl sample (Figure 3d) indicates a diagonal stripe, which, similar to the reference sample, shows enhanced C 1s intensity in the top-left and bottom-right corners (Figure 3e), which we attribute to residual MeO-2PACz remaining after delamination. However, in contrast to the reference, the stripe corresponding to the thin MeO-2PACz region also exhibits carbon-rich islands, albeit with lower overall intensity, similar to the CsCl sample with 5 nm MHP (Figure 2f). This is in stark contrast to the reference film, where the C 1s signal in this region was minimal. The I $3d_{5/2}$ distribution (Figure 3f) is anti-correlated with C 1s, as also observed in the reference film (Figure 3c). XPS spectra are shown in Figure S 16 and discussed in Supplementary Note 11. XPEEM images for additional core levels are shown in Figure S 17.

Overall, the buried interface composition matches trends seen in the 5 nm MHP case, confirming that initial film growth determines the interface chemistry. In the reference sample, thin MeO-2PACz regions remain carbon-poor, while CsCl promotes carbon incorporation across the interface.

After identifying compositional differences in the as-deposited buried interface, we examined how these variations affect the morphology and optoelectronic properties of the buried interface in annealed films, as annealing can alter the morphology, composition and phases present on the buried interface. Figure 3g presents a top-view SEM image of the buried interface of the reference sample, revealing small apparent grains and bright features that are more pronounced along a diagonal stripe, tentatively assigned to a region with thin MeO-2PACz.

To probe the radiative recombination and emission characteristics of the buried interface, cathodoluminescence (CL) measurements were performed using a set of 50 nm bandwidth band-pass filters in the 500–800 nm range. For the reference sample, emission occurred predominantly at 500 ± 50 nm, with no detectable signal at 750 ± 50 nm, the expected wavelength range for the MHP used in this study. Figure 3h and i show the 500 ± 50 nm emission for thin and thick MeO-2PACz regions respectively, highlighting that the bright features emit in this range. We attribute this emission to $PbI_2$, which has a band gap near 520 nm.[34] The increased fraction of $PbI_2$ in the thin MeO-2PACz region aligns with the as-deposited buried interface compositional analysis showing poor carbon incorporation, and with literature reports that ethanol washing of carbazole-based HTLs hampers organic precursor uptake,[24,31] as we expect washing to remove unbound MeO-2PACz molecules, forming a MeO-2PACz layer resembling a monolayer, similar to thin MeO-2PACz layer shown here.

The variation in $PbI_2$ content at the buried interface between the two regions indicates a difference in A-site cation incorporation. For the 5 nm thick reference MHP film, $Cs^+$ was distributed homogeneously (Figure S 9e), suggesting that the disparity arises from differences in organic precursor uptake between the thin and thick MeO-2PACz layers. This implies that the carbon signal (Figure 2b) can, at least in part, be attributed to the organic precursor. Interfacial $PbI_2$ can introduce defect states that quench MHP emission,[35,36] explaining the absence of 750 ± 50 nm emission despite the MHP phase being present (see XRD pattern, Figure 2l). In solar cell devices, interfacial PbI$_2$ can act as a charge extraction barrier at the buried interface due to non-optimal band alignment,[37,38] thereby decreasing PSC efficiency. In addition, PbI$_2$ has a negative influence on PSC stability as it can undergo photodecomposition, forming Pb(0) and I$_2$.[39,40] In contrast, the SEM image of the buried interface of the sample with CsCl seed shows larger apparent grains, which is favorable given that grain boundaries are associated with increased recombination due to high defect density.[41–43]

The CsCl sample showed no clear distinction between thin and thick MeO-2PACz regions (Figure 3j), therefore, a dark feature consistent with a MeO-2PACz agglomerate was identified and imaged. Based on prior topography data (Figure 1a), this location likely includes both thin and thick MeO-2PACz areas as the agglomerates form on the border between the two regions. The strongest CL signal was observed at 750 ± 50 nm (Figure 3k), with no detectable emission at 500 ± 50 nm. This indicates that CsCl suppresses interfacial PbI$_2$ formation, thereby improving the quality of the buried interface and making MHP growth during co-evaporation more robust against inhomogeneities in MeO-2PACz coverage. Figure 3l illustrates MHP formation on samples with inhomogeneous MeO-2PACz coverage.

## 2.5  Film growth on fully textured silicon bottom cells

After determining the influence of MeO-2PACz inhomogeneities on film growth on a planar substrate, we extend our analysis to textured substrates. Figure 4a shows an SEM image of a textured substrate spin-coated with a 4 mM MeO-2PACz solution. The valleys of the textured substrate appear dark compared to the pyramids, which we attribute to the accumulation of MeO-2PACz (c.f. image of substrate prior to depositing MeO-2PACz, Figure S 18a). This likely happens because the solution gets trapped in the valleys during the spin-coating process. In addition, it is possible to discern smaller MeO-2PACz aggregates on the sides of the pyramids. Even though, in an effort to prevent the accumulation of MeO-2PACz in the valleys, we reduced the MeO-2PACz solution concentration to 1 mM (Figure S 18b), and further added DMF co-solvent (Figure S 18c), which was shown to prevent the formation of micelles and lead to more homogeneous substrate coverage,[16] the MeO-2PACz agglomerates were still discernible in the valleys. A possible mitigation strategy would be vacuum-depositing MeO-2PACz; however, vacuum-deposited SAMs are still not on par with solution-processed in terms of solar cell device efficiency.

Next, using the delamination procedure, we analyzed the buried interface of MHP films grown on textured substrates to assess the influence of MeO-2PACz agglomeration on film formation. In the reference sample (Figure 4b), larger apparent grains are observed in the valley regions, while smaller, brighter grains are present along the pyramid sidewalls. These finer grains likely originate from residual PbI$_2$ at the interface. Interestingly, for MHPs deposited on textured substrates by hybrid methods, where an inorganic scaffold is first deposited, followed by the deposition of a solution of organic precursors, residual PbI$_2$ typically accumulates in the valleys, in contrast to what we observe here.[26,37,44] This happens because the organic component must diffuse over longer distances to reach the valleys. In co-evaporation, however, all precursors are deposited simultaneously, eliminating the need for long-range interdiffusion of the organic species. Still, surface modifications can strongly influence the local incorporation of the organic precursor. In our case, the accumulation of MeO-2PACz in the valleys promotes local incorporation of FA$^+$.

For the sample with a CsCl seed layer (Figure 4c), no clear differences between valleys and pyramid tops are observed. Large apparent grains cover the entire buried interface, suggesting a more homogeneous film formation. Thus, similar to our findings on planar substrates, the CsCl seed layer improves the buried interface quality by suppressing residual PbI$_2$ formation and promoting the growth of large grains across the surface.

Building on these findings, we investigated how a CsCl seed layer affects the overall performance of PSTs. Figure 4d shows a schematic of the PST device stack. We introduced ethane-1,2-diammonium iodide (EDAI) as a surface treatment for the MHP film as it can be deposited by vacuum evaporation, leading to an improvement of $V_{OC}$ in solar cell devices.[45,46] Figure S 19a shows PL of the tandem device stack without front contact, i.e. textured silicon bottom cell/IMO:H/MeO-2PACz[/CsCl]/MHP with and without EDAI treatment, indicating an increase in PLQY and thereby decrease in radiative recombination for films with EDAI. Figures 4e–h present the statistical distribution of PCE, $J_{SC}$, FF, and $V_{OC}$ values for reference devices and those with a CsCl seed layer, based on six independent MHP deposition runs. Best in-house measured PST used CsCl seed layer and reached PCE of 30.3% with $J_{SC}$ 20.3 mA cm$^{-2}$, FF 80.3%, and $V_{OC}$ 1.86 V, while certified cell reached PCE of 29.7% with $J_{SC}$ 19.8 mA cm$^{-2}$, FF 80.9%, and $V_{OC}$ 1.86 V. Results from certification measurements are marked with a black star. Figure S 20 shows the certification sheet. The higher efficiency of the in-house measured PST is linked to better current matching, leading to higher $J_{SC}$ value.

Devices with CsCl seed layer achieve higher efficiencies with a narrower distribution compared to the references. The PCE enhancement is primarily driven by improvements in FF and $V_{OC}$, which we attribute to the absence of interfacial PbI$_2$ in the CsCl samples. Interfacial PbI$_2$ has previously been shown to degrade both FF and $V_{OC}$.[36] The reference devices also display a large spread in $J_{SC}$, which correlates with significant variation in MHP top-cell band gap, ranging from 1.65 to 1.75 eV

(Figure 4i). We link the very high $J_{SC}$ (> 21.5 mA cm$^{-2}$) observed for a few references samples to low shunt resistance, which can lead to $J_{SC}$ values higher than the integrated EQE current for the limiting cell.[47] In contrast, CsCl devices exhibit less variation in band gap, centered around the optimal 1.68 eV for PSTs. Band gap control can be linked to improved incorporation of the organic precursor and suppression of PbI$_2$ formation, leading to more well-defined film growth. These trends are consistent with the single-junction results presented in our previous work, where CsCl was shown to regulate the band gap, and with the lateral variations in the MHP film growth evidenced by XPEEM in the present work. The large spread in the band gap values of the reference devices can be linked to a different degree of organic precursor incorporation.

Figure S 19b compares the EQE of wide- and narrow-band-gap reference devices, illustrating the strong impact of band gap on integrated $J_{SC}$. Figure 4j shows the EQE spectrum of the certified CsCl device, showing that the tandem is top-cell limited. This indicates that further optimization of the MHP film thickness could increase efficiency. Finally, Figure 4k presents the JV curve of the certified CsCl device, with the inset showing MPP tracking with a stabilized efficiency of 29.7%. This value represents the highest reported efficiency for PSTs featuring a fully vacuum-processed MHP absorber layer. Figure S 1a shows the evolution of PST efficiencies over time, including the present work, and highlights the deposition methods used. This comparison demonstrates that our device is the most efficient TSC employing a fully vacuum-processed MHP layer. Figure S 1b includes only PSTs in which the MHP layer was deposited using an upscalable method, excluding any process involving a spin-coating step. The figure illustrates that only a small number of PSTs employ upscalable deposition techniques, and even fewer are fabricated on industrially relevant textured silicon bottom cells with micron-scale features, with the results presented here showing one of the highest published PCE to date in this domain.

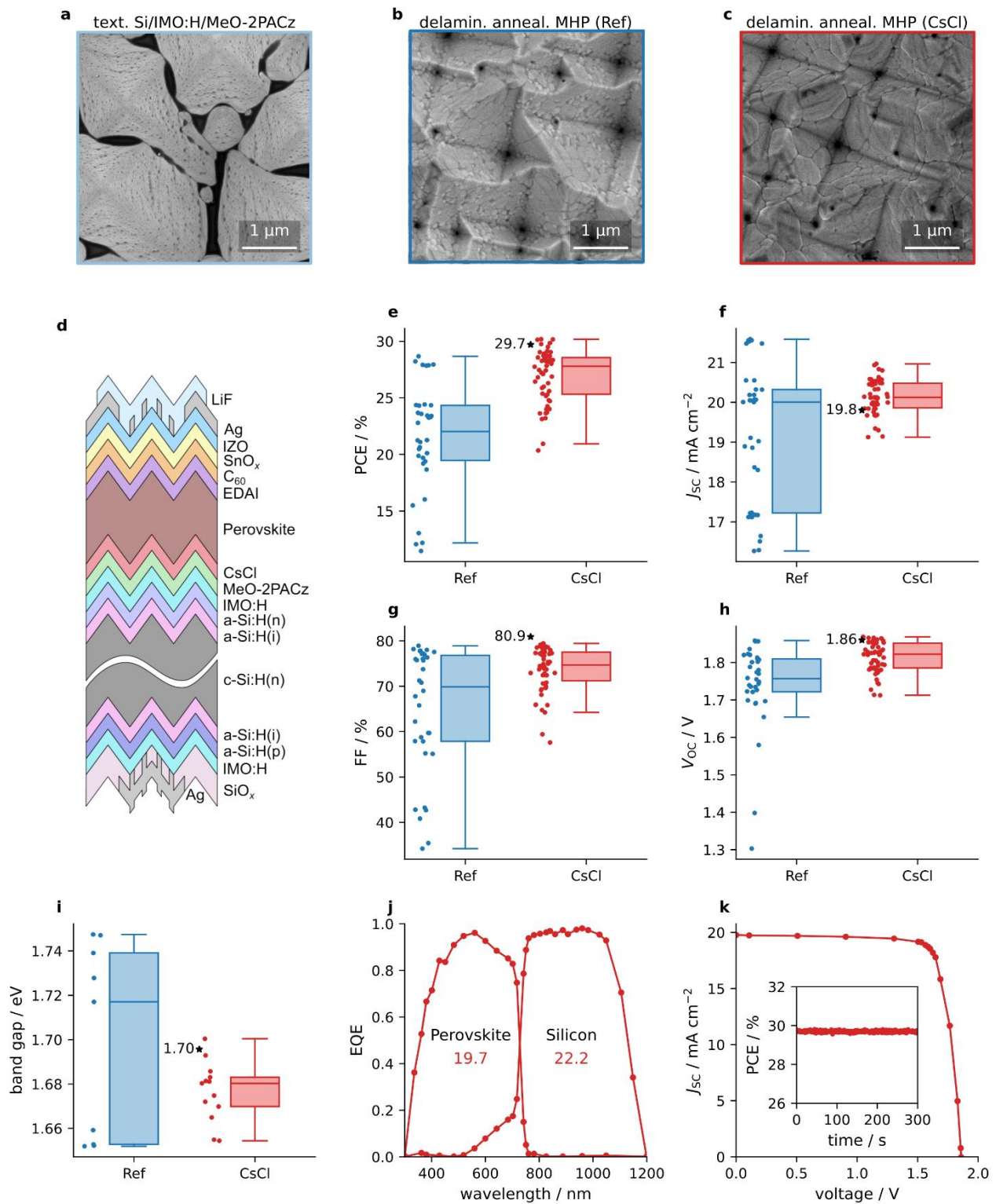

**Figure 4**: CsCl improves PCE, FF and $V_{OC}$, and reduces variation in fully textured PST top cell band gap. **a-c** Top-view SEM images. **a**, Spin-coated MeO-2PACz on a textured silicon bottom cell. **b,c**, Buried interface of delaminated reference (**b**) and MHP film with CsCl (**c**) grown on textured substrate. **d**, Schematic of the PST device stack. **e-i**, Statistical distribution of PCE (**e**), $J_{SC}$ (**f**), FF (**g**), $V_{OC}$ (**h**) and band gap (**i**) from six independent deposition runs. Black stars mark results from certification measurement. **j**, EQE spectrum of certified device. **k**, $JV$ curve of the certified device. Inset shows MPP tracking. Data shown in j and k was measured at the European Solar Test Installation (ESTI).

# 3. Summary

Using infrared scattering-type scanning near-field optical microscopy (IR s-SNOM), x-ray photoemission electron microscopy (XPEEM) and cathodoluminescence (CL), we show that incorporation of the organic precursor on the reference substrate is hindered in regions with a monolayer of MeO-2PACz, leading to the formation of interfacial $PbI_2$. In contrast, the CsCl seed layer mitigates this effect by enabling organic precursor incorporation even where MeO-2PACz is sparse. As a result, the buried interface becomes more uniform, with CsCl-treated films displaying large apparent grains and no detectable interfacial $PbI_2$ after annealing. This effect is particularly beneficial for industrially-relevant textured substrates, where MeO-2PACz tends to accumulate in the valleys of the random pyramid texture, leaving the facets comparatively uncovered. The improvement of the buried interface using CsCl leads to higher FF and $V_{OC}$ in PST devices, along with a narrower band gap distribution compared to reference PSTs. The best perovskite/silicon tandem solar cell reached an efficiency of 30.3% (29.7% certified) fabricated with a CsCl seed layer, the highest value reported to date for devices employing a fully evaporated MHP absorber layer.

# Acknowledgements

The authors acknowledge the HyPerCells joint Research School between HZB and University of Potsdam, Helmholtz Association for funding within the HySPRINT Innovation lab and the Zeitenwende project, the German Federal Ministry for Economic Affairs and Energy (BMWE) through the SHAPE project (03EE1123C), and through the program "zukunft.niedersachsen" financed by the Ministry of Science and Culture in Lower Saxony. The authors thank Helmholtz-Zentrum Berlin für Materialien und Energie for the s-SNOM measurements carried out at the IRIS beamline of BESSY II, HZB (proposal 252-13485) and acknowledge the funding by the German Federal Ministry for Education and Research (BMBF) project 05K19KH1 (SyMS). The LEEM/XPEEM measurements were carried out at the SMART microscope at the UE49-PGM beamline at the BESSY II electron storage ring operated by the Helmholtz-Zentrum Berlin für Materialien und Energie (proposals 242-12883 and 251-13156). This work was funded by the German Federal Ministry of Education and Research (Bundesministerium für Bildung und Forschung, BMBF) under Grant No. 03EW0015B (CatLab). L.d.S.C. was funded by the Deutsche Forschungsgemeinschaft (DFG, German Research Foundation) under Germany´s Excellence Strategy – EXC 2008 – 390540038 – UniSysCat.

# Author Contributions

V.Š. planned the experiments, coordinated the work, prepared the samples for and conducted or aided in s-SNOM, XRD, CL, and XPEEM measurements, and prepared the figures and manuscript. S.S. Led the preparation, optimization and analysis of perovskite–silicon tandem solar cells. A.M. prepared perovskite–silicon tandem solar cells and performed SEM. S.W. and D.A.-R. carried out and interpreted the CL measurements. M.J.P., L.C.T., J.-W.H., M.-A.M., L.d.S.C., and T.S. performed and interpreted the XPEEM measurements. R.F.D. and S.M aided in XPEEM measurement preparation. E.W. and R.M. measured and interpreted XRD data. A.V., S.D., and L.P. performed and analyzed s-SNOM measurements. M.L. optimized the EDAI passivation layer. A.H., J.K., P.W. and B.S. optimized the silicon bottom cell. V.Š. and L.K. discussed extensively on data analysis and the overall interpretation of XPEEM and s-SNOM data. L.K., S.A., and M.R. supervised the project and contributed to its planning. All authors contributed to the writing of the manuscript.

# Methods

## Materials

Lead iodide (99.99%, trace metals basis) from TCI, Lead bromide from TCI, Cesium iodide from abcr, Formamidinium iodide (FAI) from Great Cell Solar Materials were used for MHP deposition. Cesium chloride from abcr was used for deposition of the seed layers. MeO-2PACz from TCI and ethanol from VWR Chemicals were used for hole selective contacts. Ethylenediammonium diiodide (EDAI) from Sigma Aldrich was used for MHP surface passivation. $C_{60}$ (sublimed) from CreaPhys GmbH and BCP from Sigma Aldrich were used for electron-selective contact. Ag pellets from thermo scientific were used for top electrode. LiF from Sigma-Aldrich was used for the antireflective coating. All chemicals were used without further purification.

## MHP co-evaporation

The perovskite layer was deposited via co-evaporation using $PbI_2$, $PbBr_2$, CsI, and FAI as precursors, employing the CreaPhys "PEROvap" thermal evaporation system located inside a nitrogen glovebox. Deposition rates were continuously monitored with quartz crystal microbalance (QCM) sensors, and the source temperatures were dynamically adjusted based on the measured rates to maintain stable deposition conditions. Source preheating began at a chamber pressure of 2 nbar. During deposition, the inner shielding temperature was maintained at −25 °C, the substrate temperature at 20 °C, and the substrate holder rotated at 10 rpm. The deposition process lasted for 94 minutes for a 500 nm thick film. Thickness was varied during a deposition using a wedge shutter. Where stated that samples were annealed, the samples were annealed at 180 °C for 5 minutes in a nitrogen atmosphere after the deposition. Otherwise, the samples were not annealed. MHP films used in perovskite-silicon tandem solar cells were annealed.

## Monolithic perovskite/silicon tandem solar cell preparation

### Bottom cell

For fully evaporated PSTs, double-sided textured silicon bottom cells are used. The bottom cells are based on 260 µm thick n-type (100) float-zone c-Si wafers, which are chemically etched on both sides to form a random pyramid texture with an average height of ~2 µm. After chemical etching and standard RCA cleaning, doped and intrinsic amorphous silicon layers are deposited by plasma-enhanced chemical vapor deposition (PECVD) on both sides of the textured wafer to form a silicon heterojunction (SHJ). Depositions are performed in a PECVD cluster tool (INDEOtec SA) operating at 13.56 MHz.

On the rear side of the silicon bottom cell, a 5 nm intrinsic (i) and 7 nm p-doped (p) a-Si:H layers are deposited by PECVD. The back contact of the bottom cell, and later the tandem cell, consists of a 30 nm sputtered metal-doped $In_2O_3$-based transparent conducting oxide (TCO) layer (IMO:H) with a 1 $cm^2$ area and a screen-printed silver grid. This TCO/silver stack defines the active area of the device and serves as the back contact. Additionally, a 180 nm a-SiO$_x$(i) layer is deposited by PECVD to act as a rear reflector, minimizing parasitic absorption in the silver contact[48] and providing scratch protection during top-cell processing. Finally, a 400 nm silver patch (1 $cm^2$) is sputtered on top.

On the front side of the silicon bottom cell, a stack of 5 nm intrinsic a-Si:H(i) and 5 nm n-doped a-Si:H(n) is deposited by PECVD. A nominal 20 nm sputtered indium-based TCO is then added on top of this stack as the recombination contact. This contact, also 1 $cm^2$ in area, aligns with both the rear TCO/silver contact and the front contact of the subsequently processed MHP top cell.

### Top cell

Before processing the MHP top cell, the silicon bottom cell is dynamically spin-coated with ethanol to remove dust particles, followed by a 15 min ozone cleaning. The HTL is then deposited by spin-coating 90 µL of 4 mM MeO-2PACz solution, followed by annealing at 100 °C for 10 min. A 20 nm thick CsCl seed layer is subsequently deposited by thermal evaporation in vacuum using an OPTIvap evaporator (Creaphys/M. Braun Inertgas-Systeme GmbH).

Following MHP evaporation and annealing, a 1 nm EDAI passivation layer and a 16 nm $C_{60}$ electron-transporting layer are deposited by thermal evaporation in the same vacuum chamber as the seed layer. The top contact stack is completed with a 20 nm SnO$_x$ layer deposited via ALD (using an ALD system from Arriadance LLC), a 40 nm sputtered indium zinc oxide (IZO) layer (using a sputtering tool from FHR Anlagenbau GmbH), and a 500 nm evaporated silver frame with grid fingers (using an evaporation chamber from CREAVAC PVD AG). As last step a 100 nm thick lithium fluoride layer (LiF) was deposited as anti-reflection coating by thermal evaporation.

### Scanning electron microscopy and cathodoluminescence

A Zeiss MERLIN field emission scanning electron microscope with a GEMINI II optical column from Zeiss was used for measuring scanning electron microscopy (SEM) images. An accelerating voltage of 5 kV and a current of 100 pA were used. For CL imaging, scanning electron microscopy

(SEM) images were acquired using a Zeiss MERLIN field emission SEM scanning electron microscope equipped with a GEMINI II electron optical column and a DELMIC Sparc CL system. The SEM imaging and CL measurements were performed at an accelerating voltage of 5 kV and a probe current of 50 pA.

### Steady-state photoluminescence

A homebuilt setup with a power meter was used in a $N_2$ glovebox. The PL was collected using a fiber connected to a CCD-array QE Pro spectrometer (Ocean Optics). The spot size on the sample is 0.14 $cm^2$. The samples were excited with a continuous-wave laser at 520 nm emission wavelength and an incident power of 3.6 mW on the sample. The samples were illuminated from the glass side. An average of 10 consecutive measurements is calculated, with an integration time of 200 ms.

### X-ray diffraction

The XRD patterns are derived from binned two-dimensional grazing-incidence wide-angle x-ray scattering (GIWAXS) images. GIWAXS measurements were carried out using a high-flux laboratory system at the X-ray Corelab, Helmholtz-Zentrum Berlin, featuring an Excillum liquid metal-jet source with a 68:22:10 Ga:In:Sn (weight fractions) alloy anode operated at 70 kV. The setup uses the Ga K-L3 and K-L2 emission line (9.2517 keV and 9.2248 keV) as the primary x-ray energy. x-rays were focused by multilayer optics with a nominal focus length of 2500 mm through a vacuum flight tube onto the sample, producing a horizontal spot size of approximately 1 mm at an incident angle of 8°. The multilayer optics filters out other emission lines and the bremsstrahlung background. Diffraction patterns were collected using a Dectris PILATUS3 S 1M 2D detector.[49] Measured data was processed and binned using INSIGHT Python package.[50]

### Photoluminescence spectroscopy

A homebuilt setup with a power meter was used in a $N_2$ glovebox. The PL was collected using a fiber connected to a CCD-array QE Pro spectrometer (Ocean Optics). The spot size on the sample is 0.14 $cm^2$. The samples were excited with a continuous-wave laser at 520 nm emission wavelength and an incident power of 3.6 mW on the sample. The samples were illuminated from the glass side. An average of 10 consecutive measurements is calculated, with an integration time of 200 ms.

# Low-energy electron microscopy and x-ray photoelectron emission microscopy

XPEEM and low-energy electron microscopy (LEEM) measurements were performed using the aberration-corrected and energy-filtered SMART spectro-microscope located at the UE49-PGM undulator beamline at the BESSY II synchrotron radiation facility (Helmholtz-Zentrum Berlin, HZB). In XPEEM mode, the instrument provides an energy resolution of 180 meV and a lateral spatial resolution of 18 nm. LEEM mode enables imaging with a lateral resolution of 2.6 nm.[51] All measurements were performed with a field of view of 12.4 μm. The samples were kept in nitrogen atmosphere during transport, mounting onto the sample holders, and loading into the load lock chamber. After pump-down to UHV, they were transferred immediately to the analysis chamber for measurements.

XPEEM measurements were conducted at fixed photon energies which were chosen such that the photoelectrons of the selected core-level peak appear close to 150 eV. This approach ensured a consistent probing depth across different elemental spectra by maintaining similar photoelectron kinetic energies.

If elastic-scattering effects are neglected, the information depth (ID) can be estimated as:[52]

$$ID = \lambda_{in}\, cos(\alpha)\, ln\left[\frac{1}{1-(P/100)}\right]$$

Where $\lambda_{in}$ is the inelastic mean free path, $\alpha$ is the angle of emission and $P$ is the specified percentage of detected signal. $\alpha$ is 90° in the SMART spectro-microscope. $\lambda_{in}$ is 0.75 nm for elemental carbon[53] and 0.72 for elemental lead.[54] For $P = 95\%$, we obtain 2.2 nm for carbon as well as lead, as a rough estimate of the ID in MeO-2PACz and MHP respectively.

The following photon energies were used for specific elemental core levels:

- 772 eV – I $3d_{5/2}$
- 437 eV – C 1s
- 225 eV – Br 3d
- 295 eV – Pb $4f_{7/2}$
- 351 eV – Cl 2p
- 600 eV – In $3d_{5/2}$
- 680 eV – O 1s

- 882 eV – Cs 3$d_{5/2}$
- 286 eV – P 2p and secondary electron cutoff measurement

LEEM imaging was performed by acquiring a single image with an integration time of 1 s. Electron energy that gave best contrast was chosen based on an electron energy scan in range from -4 to 10 eV. schematically shows XPEEM data processing. XPEEM was initially acquired as a kinetic energy scan (Figure S 21a), with 1 s integration per image and a step size of 0.25 eV. The XPS spectrum obtained as spatial average of individual kinetic scan images was used to identify the peak maximum (or maxima, where applicable) and an appropriate baseline energy (Figure S 21b). The final three energy-scan images near the baseline were averaged and subtracted as background correction of XPS spectra. High-quality images were then recorded at both the peak maximum and baseline energy (Figure S 21c). For each condition, five images were collected with an integration time of 10 s per image. High-quality images were first averaged to enhance contrast (not shown on the schematic). Elemental contrast was then obtained by subtracting the averaged baseline image from the averaged peak maximum image (Figure S 21d).

## Scattering-type scanning near-field optical microscopy

IR s-SNOM measurements were performed at the IR-nanospectroscopy end station of the IRIS beamline at the BESSY II synchrotron facility using a commercial s-SNOM nanoscope (NeaSNOM, Neaspec Attocube systems, Haar, Munich, Germany) and broadband synchrotron source.[55,56] The measurements were carried out at room temperature, with the s-SNOM setup continuously purged using dry nitrogen to minimize the influence of atmospheric moisture and $CO_2$.

The measurements were conducted in tapping mode, with a setpoint of 70% and a typical contact amplitude of 55 nm. AFM probes coated with platinum–iridium (Pt–Ir) (Arrow™ NCPt, NanoWorld, Neuchâtel, Switzerland) were used. These probes feature a nominal tip radius of <25 nm. The cantilever was driven close to its resonance oscillation frequency of approximately 255 kHz.

The optical signal demodulated at the second harmonic of the AFM oscillation frequency is used in this study. The nano-IR images were collected in zero-path difference heterodyne imaging (known as "white light" imaging) mode using the broadband infrared synchrotron radiation.[57]

Nano-IR spectra were acquired by locating the AFM tip at defined sample positions and recording the interferograms. The interferometric scan length was 700 μm, measured in 768 pixels, with integration time of 100 ms per pixel. The nominal spectral resolution at the parameters used in the measurement is 7.2 cm$^{-1}$. For each point of interest 10 interferograms were averaged to improve the signal to noise ratio.

Raw interferograms were Fourier-transformed using a custom in-house script developed at the IRIS beamline using SciLab (open-source software). The data were apodized using an asymmetric window based on the three-term Blackman–Harris function and subjected to zero-filling with a factor of four. Infrared amplitude and phase spectra were extracted from the processed interferograms. All spectra were referenced against a clean Si/IMO:H substrate to account for instrumental and environmental background signal, a linear baseline was subtracted from the resulting spectra. The reference and baseline-corrected optical phase shift signals are shown as nano-IR spectra.

AFM images were pre-processed using the open-source software Gwyddion.[58] Plane subtraction was applied to correct for surface tilt, and median of difference for row alignment within the thin MeO-2PACz region to minimize imaging artifacts prior to data analysis.

### Perovskite film delamination

BLUFIXX, a UV-curable adhesive, was used to delaminate samples by enabling controlled mechanical separation after rapid curing under UV light. For XPEEM and CL measurements, prior to delamination, a 25 nm layer of ITO was sputtered onto the surface to ensure efficient charge extraction.

### Current-voltage measurements

The perovskite/silicon tandem solar cells were measured in air under AM1.5G spectrum with an intensity of 1000 W/m² using a Wavelabs Sinus-70 LED class AAA solar simulator. For the correct illumination settings, the spectra of all 21 LEDs were measured from 0 % to 100 % in 10 %-steps. Additional, the spectral responses of both sub cells are required (EQE measurement) which are uploaded to a self-written MATLAB code which generates a spectrum from the measured single-LED spectra. This spectrum fulfils two conditions. First, the current generated under AM1.5G and under the solar simulator spectrum is the same simultaneously for both sub cells. Second, the solar simulator spectrum needs to be as close as possible to the AM1.5G spectrum. Therefore only two LED channels, one in the blue light regime and one from the red light regime, are slightly changed in their intensity. After finding the LED settings, the current generated in a RG780-filtered WPVS reference cell (calibrated by Fraunhofer ISE CalLab) was calculated and used for calibrating the illumination intensity of all 21 LEDs of the AM1.5G spectrum. The back contact of the tandem solar cell was contacted with a metal vacuum chuck (with controlled temperature, fixed at 25°C) and the front side was contacted with two Au probes (source and sense of a 4-wire-setup) on the deposited silver contact. While measuring a 1 cm² shadow mask was used to illuminate the active area only.

The JV scans were recorded with a LabView software using a step size of 0.02 V from -0.1 V to 2 V, 20 ms integration time and 40 ms settling time.

### External Quantum Efficiency Measurements

The external quantum efficiency (EQE) of each sub-cell of the tandem solar cell device was measured with a home-built setup by using chopped monochromatic light from a Xe and He lamp (79 Hz) which illuminate almost the full 1 cm² active area. For measuring the perovskite top cell, the silicon bottom cell was saturated using a NIR LED with emission at 850 nm and applying a bias voltage of 0.6 V to maintain short-circuit conditions. For the silicon bottom cell, the perovskite top cell was saturated with blue light using a LED emission at 455 nm and a bias voltage of 0.9 V.

# Supplementary information

**Supplementary Note 1**: Differences between IR s-SNOM spectra of thin and thick MeO-2PACz layers

The reduced intensity of bands at 1030 cm$^{-1}$ (O-H bend) and 1290 cm$^{-1}$ (O-H bend and P=O stretch) in the thin region suggests reduced hydrogen bonding between MeO-2PACz molecules and indicates binding to the substrate.[28] Additionally, the absence of the 800 cm$^{-1}$ band associated with the out-of-plane γ(C−H) vibration of the carbazole group in the thin MeO-2PACz spectrum indicates a preferential orientation of the carbazole moiety with respect to the ITO substrate, confirming the formation of a self-assembled monolayer.[28] Additional decreased absorption was detected at 1494 cm$^{-1}$, 1214 cm$^{-1}$, and 1065 cm$^{-1}$. Since these bands are not directly associated with molecular orientation or hydrogen bonding,[28] the observed decrease cannot be directly explained and comprehensive analysis falls outside the scope of this work. A notable reduction is also seen in the band at 1160 cm$^{-1}$, whose assignment in the literature is ambiguous: it has been linked both to P=O stretching[29,30] and to CH$_2$ bending and carbazole ring vibrations[28]. We attribute the decrease to partial tridentate binding of MeO-2PACz forming P–O–In bonds, which selectively decreases the P=O contribution. The remaining intensity can originate either from CH$_2$ bending and carbazole ring modes, or from unbound P=O groups.

**Supplementary Note 2**: MeO-2PACz molecular layer thickness.

The reported thickness of a washed 2PACz film, which likely corresponds to a monolayer, is ~0.8 nm.[59] Since the only structural difference between 2PACz and MeO-2PACz is the methoxy substituent on the carbazole ring, which is not expected to significantly affect molecular length, we take 0.8 nm as a rough estimate for the MeO-2PACz monolayer thickness. The thickness difference of 6 nm would therefore correspond to several molecular layers.

**Supplementary Note 3**: Potential influence of MeO-2PACz layer inhomogeneity on solar cell device performance.

From a device perspective, we expect that solar cells would not be severely impacted by observed MeO-2PACz inhomogeneities, since full coverage of the interface should still prevent the enhanced interface recombination or possible shunt pathways that might arise from incomplete SAM coverage.[60,61] While localized agglomerates could in principle lead to optical losses or hindered charge extraction, these features cover only a small fraction of the overall interface and are thus unlikely to pose a major limitation in device operation. However, we later show that such inhomogeneities influence MHP formation during co-evaporation, and as the fractions of the substrate covered by thick and thin MeO-2PACz layers would likely be difficult to control, this would lead to variations in MHP composition and consequently poor solar cell device reproducibility. This could explain the poor reproducibility observed for co-evaporated devices without a seed layer shown in our earlier work.[32] Herein, we show that film formation is more homogeneous when using a CsCl seed layer, explaining the reduced variability in film properties and solar cell device performance.

**Supplementary Note 4**: MeO-2PACz overlayer thickness estimate from XPEEM measurements

For energy of 150 eV and MeO-2PACz overlayer material, the estimated ID for 95% of the signal is ~2.2 nm (for details see Methods). This is consistent with nano-IR spectra showing that the thin MeO-2PACz layer resembles a monolayer, as we roughly estimate the MeO-2PACz monolayer thickness to be ~0.8 nm based on literature values for 2PACz (see discussion above).[59]

Assuming the thick MeO-2PACz region is ~6 nm thicker than the thin one, based on the height profile in Figure 1b, we can estimate the attenuation of the In $3d_{5/2}$ signal by MeO-2PACz layer using a simple exponential decay model[52]:

$$I(d) = I_0 \exp\frac{-d}{\lambda_{in}}$$

Where $I(d)$ is the photoelectron intensity after passing through an overlayer of thickness $d$, $I_0$ is the initial photoelectron intensity and $\lambda_{in}$ is the inelastic mean free path. At 150 eV kinetic energy, $\lambda_{in} \approx$ 0.75 nm for elemental carbon,[53] which we use as a rough estimate for MeO-2PACz. For an overlayer thickness of $d = 6$ nm this yields $I(d) \approx 0.00034\ I_0$, meaning only 0.034% of the signal is transmitted. This explains the strong contrast in In $3d_{5/2}$ signal between thin and thick MeO-2PACz regions.

**Supplementary Note 5**: Additional XPEEM analysis of Si/IMO:H/MeO-2PACz sample

The secondary electron cutoff spectra (Figure S 4a) obtained by averaging the signal from two regions with different MeO-2PACz film thicknesses highlighted by green and orange circles on the LEEM image in the inset of Figure S 4a, reveal that the region with a thin SAM layer has a work function of 4.89 eV, 0.16 eV lower than the region with a thick SAM layer (5.05 eV). This goes in line with a previously reported increase in work function with increasing thickness of spray-coated 2PACz molecules, which closely resemble MeO-2PACz. For 2PACz, the work function increased from 4.62 eV to 5.02 eV depending on the number of spray-coating cycles. This effect was assigned to the interface dipole due to molecular assembly on the substrate surface.[59] A more subtle increase is to be expected for MeO-2PACz due to a lower dipole moment.[12]

**Figure S 4**b-e shows core-level XPS spectra obtained by averaging the signal intensity in the regions assigned to thin and thick MeO-2PACz. Note that for the P 2p core level the two areas could not be clearly distinguished on the images due to low signal intensity, therefore the whole field of view had to be considered.

The C 1s core level XPS spectrum (Figure S 4b) shows that the peak within the region with thin MeO-2PACz has a similar shape and peak maximum position as the peak within the region with thick MeO-2PACz layer, suggesting that the carbon signal in the region with lower signal intensity indeed originates from remaining MeO-2PACz molecules, in line with the IR s-SNOM results.

The O 1s core level spectrum (Figure S 4d) exhibits distinct peak shapes in regions with thin and thick MeO-2PACz. In the thin MeO-2PACz region, the O 1s peak has a maximum around 530.5 eV, with additional contributions at higher binding energy. This shape is consistent with spectra observed for bare indium–tin oxide (ITO) and ITO coated with a very thin MeO-2PACz layer.[12,59] The O 1s image at 530.5 eV (Figure S 5a) closely matches the In $3d_{5/2}$ distribution, showing stronger intensity in the thin MeO-2PACz region and confirming that this peak originates from the O–In bond. In contrast, thick MeO-2PACz regions show a maximum at 532.75 eV, attributed to either C–O–C[12] of MeO-2PACz or the P–OH bond in the case of 2PACz molecule.[59] The O 1s core level spectrum of $In_2O_3$ also has a shoulder at this energy, associated with hydroxide species and contaminants in ITO.[12,59] Since both contributions overlap at this binding energy, the O 1s image at 532.75 eV (Figure S 5b) appears mostly homogeneous, with the regions with thin MeO-2PACz layer having a slightly higher signal intensity. The P 2p image (Figure S 5c) shows a lower intensity in the region with thin MeO-2PACz which is expected due to the lower thickness of the MeO-2PACz layer. However, it is difficult to clearly distinguish this region on the image due to the low signal intensity.

**Supplementary Note 6**: Additional XPEEM analysis of Si/IMO:H/MeO-2PACz/CsCl sample

The Cl 2p (Figure S 7b) and Cs $3d_{5/2}$ (Figure S 7e) peaks show small shifts (~0.25 eV) to higher binding energy. While such shifts could arise from sample charging, the more insulating nature of MeO-2PACz compared to IMO:H[62] would suggest a shift to higher binding energy in thick MeO-2PACz regions, opposite to what is observed here. In thick MeO-2PACz regions, the O 1s spectrum (Figure S 7d) displays two peaks at 530.75 eV and 527 eV. The origin of the 527 eV peak is currently unclear and requires further investigation.

**Supplementary Note 7**: Origin of C 1s signal in Si/IMO:H/MeO-2PACz/5 nm MHP sample

With a thin MHP film atop the MeO-2PACz layer, the carbon signal can originate not only from MeO-2PACz and adventitious carbon, as discussed earlier, but also from the organic precursor FAI. Assuming that MeO-2PACz is the primary source of the carbon signal explains the C 1s distribution well, as there is a clear difference in the C 1s intensity between regions associated with the bright region in LEEM, which we therefore assign to thin MeO-2PACz layer, and the surrounding film, which we assign to thick MeO-2PACz. The observed inhomogeneity in signal intensity within the thick MeO-2PACz regions is likely due to uneven coverage by the nominally 5 nm thick MHP layer, which would unevenly attenuate the signal originating from the MeO-2PACz interlayer. The measured C 1s binding energy of ~284.5 eV is consistent with literature values for MeO-2PACz[12,22] as well as the value obtained for the reference sample without MHP (Figure 1e).

For $FA^+$, a peak is expected at higher binding energy (~288-289 eV),[22,63] characteristic of the C=N bond. As no such peak was observed (see C 1s scan for the 20 nm thick sample with CsCl layer in Figure S 10a), and high-resolution scans showed no change in contrast within the C 1s peak across different binding energies (Figure S 10b,c), high-quality images were only acquired at the binding energy corresponding to the C 1s peak maximum. In addition to $FA^+$, degradation products of FAI could also contribute to the C 1s peak. However, FAI decomposition was shown to generate hydrogen cyanide and s-triazine,[64] which are expected to have binding energies similar to or higher than $FA^+$. Therefore, MeO-2PACz remains the most plausible source of the observed C 1s signal distribution in the reference sample with 5 nm MHP.

**Supplementary Note 8**: Additional XPEEM analysis of Si/IMO:H/MeO-2PACz/5 nm MHP sample

The thin MeO-2PACz region shows lower O 1s (Figure S 9c) and P 2p (Figure S 9d) signals, which can be attributed to the lower concentration of MeO-2PACz, and a higher $Pb\ 4f_{7/2}$ signal (Figure S 9b). Given that the thin MeO-2PACz region is rich in lead and iodine and poor in carbon, we infer that lead iodide forms preferentially in this region. Br 3d (Figure S 9a) and $Cs\ 3d_{5/2}$ (Figure S 9e) signals are homogeneously distributed with no pronounced contrast.

The $Pb\ 4f_{5/2}$ XPS spectrum (Figure S 8d) exhibits a shift of approximately 0.5 eV toward lower binding energy in the thick MeO-2PACz region compared to the thin region. The origin of this shift is not fully understood. One possible explanation is coordination of $Pb^{2+}$ by a Lewis base[65] (e.g., $FA^+$ or phosphonic acid oxygen), however we would then expect to observe the appearance of two distinct components rather than a uniform shift. At present, we cannot exclude the possibility that the shift arises from measurement artifacts or subtle variations in the local environment of lead, and further systematic studies would be required to clarify its origin.

**Supplementary Note 9**: Additional XPEEM analysis of Si/IMO:H/MeO-2PACz/CsCl/5 nm MHP sample

The observed shift in the $I\ 3d_{5/2}$ (Figure S 11a) core level can be related to the $Pb\ 4f_{7/2}$ spectrum (Figure S 11d), which shows a feature at ~137 eV attributed to Pb(0), a known degradation product of MHPs under x-ray irradiation.[66–68] $Pb^{2+}$ degradation to Pb(0) is partial in the thick MeO-2PACz region and complete in the thin MeO-2PACz region. Formation of Pb(0) can also explain the observed shifts in Br 3d (Figure S 11c) and Cl 2p (Figure S 11e).

We note that the reference Si/IMO:H/MeO-2PACz/5 nm MHP sample was exposed to lower overall x-ray dose, as the XPS peak maxima for each core level were determined on fresh areas of the sample, and the XPEEM images were then acquired at binding energies set from those scans. As a result, potential degradation from prolonged x-ray exposure is likely not reflected in the corresponding XPS spectra. By contrast, for the 5 nm MHP film grown on CsCl seed layer and the other samples discussed in this chapter, the XPS spectra were acquired on the same spot prior to XPEEM imaging, increasing the chance of beam-induced degradation. Additionally, we note that the 5 nm MHP film deposited on a 10 nm CsCl seed layer likely deviates from stoichiometric composition, as evidenced by the high fraction of $Cs^+$ and $Cl^-$. The stability of such a Cs- and Cl-rich film relative to stoichiometric MHP or $PbI_2$ remains uncertain. For these reasons, a direct stability comparison between the reference and CsCl-seeded 5 nm MHP films is not possible. Still, we note the increased stability of 5 nm MHP film grown on CsCl under x-rays in regions with thick MeO-2PACz, suggesting partial stabilization of $Pb^{2+}$. This effect could arise either from interactions with unbound MeO-2PACz molecules or from differences in film formation.

The O 1s peak shape (Figure S 11g) differs from that of the CsCl sample without MHP (Figure S 7d), showing a main peak at 530.5 eV with an additional shoulder at lower binding energy. The peak shape is consistent across thin and thick MeO-2PACz regions. Figure S 12c-j shows XPEEM images of other core levels. The Pb $4f_{7/2}$ images at 138.25 eV ($Pb^{2+}$, Figure S 12c) and 137 eV (Pb(0), Figure S 12d) show that Pb(0) has a higher intensity in the region with thin MeO-2PACz. Br 3d (Figure S 12e,f) appears in islands, similar to Pb $4f_{7/2}$ and C 1s. O 1s and Cs $3d_{5/2}$ show a homogeneous distribution, while slight inhomogeneities in Cl 2p signal distribution (Figure S 12i) can be explained by the observed XPS peak shift (Figure S 11e) as only one XPEEM image at binding energy of 199 eV was measured, corresponding to the XPS peak maximum of the region with thick MeO-2PACz, which means that the $Cl^-$ in the region with thin MeO-2PACz is likely slightly underestimated.

**Supplementary Note 10**: Additional XPEEM analysis of buried interface of 100 nm thick MHP film grown on Si/IMO:H/MeO-2PACz substrate

XPS spectra of the buried interface (Figure S 14) further support these observations. Weak O 1s (Figure S 14f) and P 2p (Figure S 14h) signals are present in the thick MeO-2PACz region, indicating residual delaminated MeO-2PACz molecules, but are absent in the thin MeO-2PACz region. Conversely, a weak In $3d_{5/2}$ signal appears in the thin MeO-2PACz region (Figure S 14e), originating from the underlying ITO:H layer. We also note a shoulder characteristic of Pb(0) at ~137.25 eV in the Pb 4f$_{7/2}$ spectrum of the thin MeO-2PACz region, showing that MHP in this region is more susceptible to degradation. In addition, the C 1s peak in the thin MeO-2PACz region is shifted by ~0.5 eV to lower binding energy (Figure S 14b), suggesting a difference in carbon's chemical state, with the C 1s signal primarily originating from the organic precursor in the thin MeO-2PACz region and from delaminated MeO-2PACz molecules in the thick MeO-2PACz region. We rule out band banding and charging as similar shifts should then be observed in other core level spectra.

A high-binding-energy shoulder shifted by ~1.5 eV from the main peak is visible in the I 3d$_{5/2}$ spectrum of the thick MeO-2PACz region (Figure S 14a). Similar features have been reported by Oz and Olthof on thin (≤ 200 nm) FAPbI$_1$Br$_2$ MHPs, attributing them to either different surface iodide bonds or partial charging effects.[22] The XPEEM image acquired at the energy corresponding to the shoulder shows a homogeneous lateral distribution (Figure S 15b), making localized charging in the thick MeO-2PACz region an unlikely explanation. We therefore tentatively attribute the shoulder to different iodide surface bonds.

Additional LEEM and XPEEM images (Figure S 15) show that the Pb 4f$_{7/2}$ peak at 138.75 eV (Pb$^{2+}$, Figure S 15c) follows the same spatial distribution as I 3d$_{5/2}$, with stronger intensity in the thin MeO-2PACz region and attenuation in the thick region due to the MeO-2PACz overlayer. O 1s (Figure S 15g,h) and P 2p (Figure S 15j) distributions match that of C 1s, confirming MeO-2PACz as the dominant carbon source. The Cs 3d$_{5/2}$ signal shows slight lateral inhomogeneity in the thin MeO-2PACz region (Figure S 15i), whereas Br 3d exhibits more pronounced inhomogeneities, forming Br-rich islands (Figure S 15f).

**Supplementary Note 11**: Additional XPEEM analysis of buried interface of 100 nm thick MHP film grown on Si/IMO:H/MeO-2PACz/CsCl substrate

XPS spectra (Figure S 16) reveal that, unlike the reference, the I 3d peak does not show a high-binding-energy shoulder (Figure S 16a). However, a ~0.5 eV shift towards lower binding energy is observed for Br 3d in the thick MeO-2PACz region (Figure S 16c). As in the reference, the C 1s peak in the thin MeO-2PACz region is shifted by ~0.5 eV, suggesting the same chemical origin of the signal but with higher intensity in the CsCl sample. No Pb(0) shoulder is observed in Pb 4f$_{7/2}$, indicating improved resistance of the interface to x-ray–induced degradation. This suggests that the strong degradation previously observed for the 5 nm MHP on CsCl is specific to thin films with non-stoichiometric composition. O 1s and P 2p signals are confined to the thick MeO-2PACz region, whereas In 3d$_{5/2}$ is detectable in both regions.

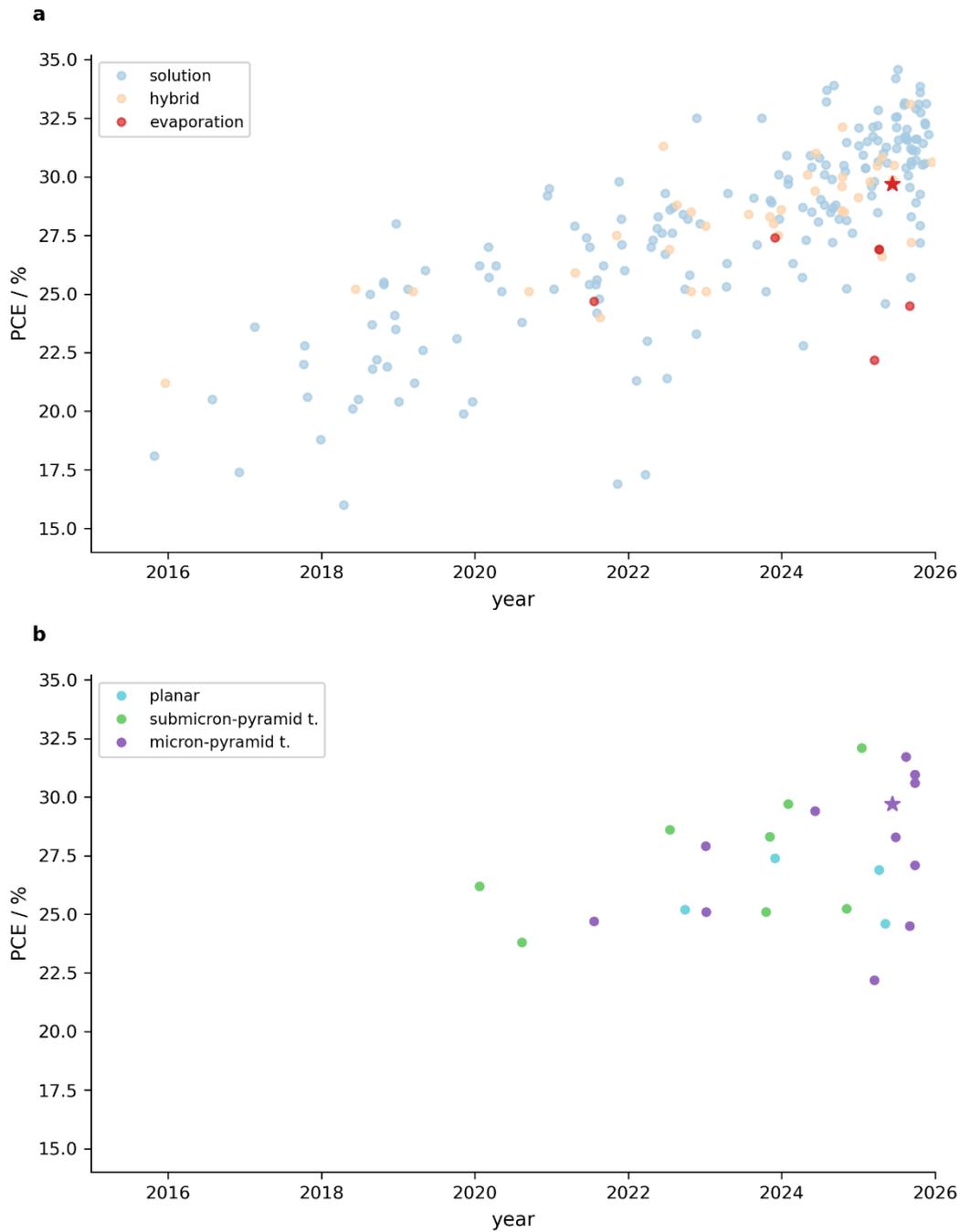

**Figure S 1**: Development of PST efficiency, categorized by deposition method (**a**) and bottom cell texture on the side facing the top cell (**b**). b shows only PSTs where MHP film was deposited using upscalable deposition techniques, i.e. excluding processes which include a spin-coating step. Results from this work are marked with a star, with the certification date as the x-axis value (10[th] of June 2025). The plot was made on 26[th] of November 2025. Data prior to May 2024 was taken from ref. 69.

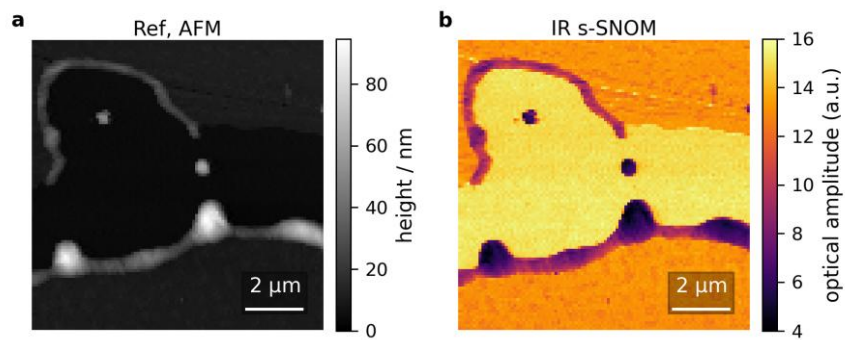

**Figure S 2**: Sample with inhomogeneous MeO-2PACz coverage. **a**, Topography image shown in Figure 1a with full height range. **b**, Near-field IR optical amplitude image in 2$^{nd}$ demodulation harmonic.

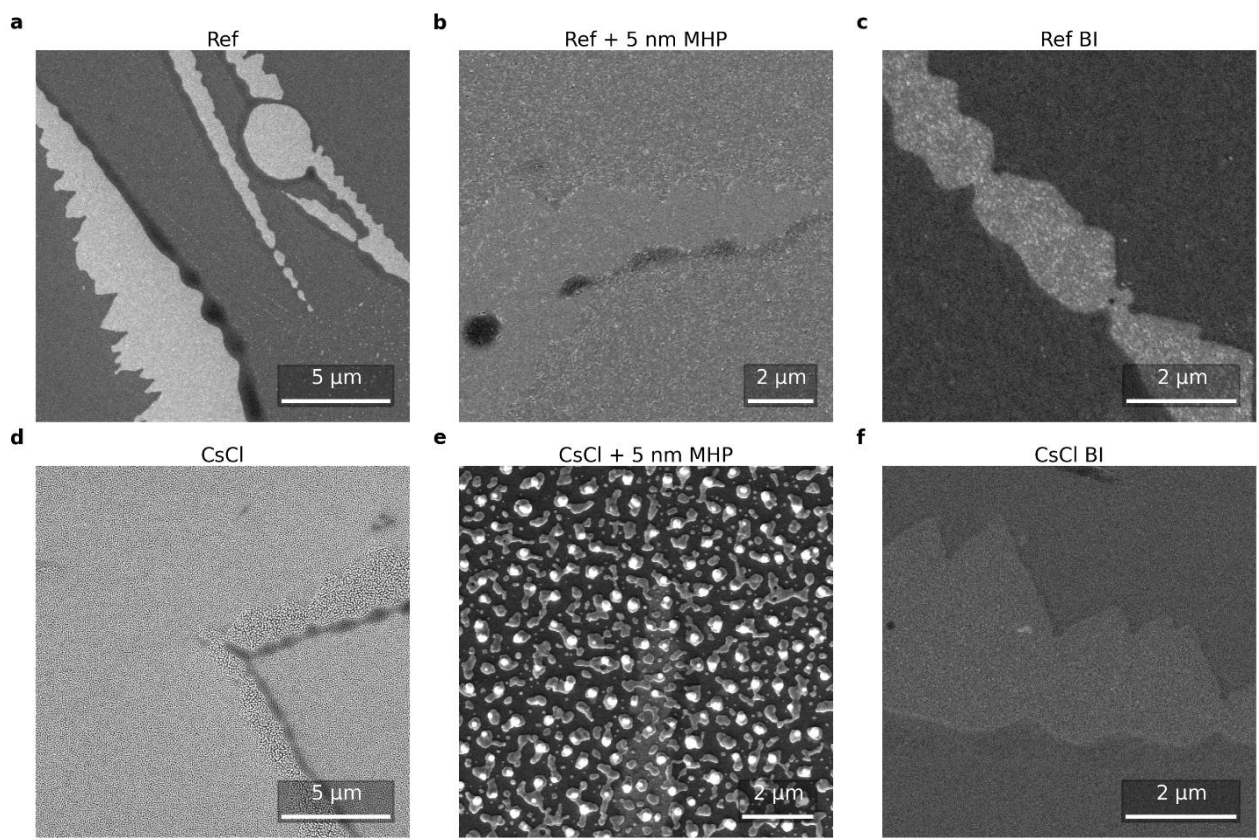

**Figure S 3**: SEM images of Si/IMO:H/MeO-2PACz (**a**), Si/IMO:H/MeO-2PACz with 5 nm MHP (**b**), buried interface of 100 nm thick MHP (**c**), Si/IMO:H/MeO-2PACz/CsCl (**d**) Si/IMO:H/MeO-2PACz/CsCl with 5 nm MHP (**e**), 100 nm thick MHP grown on CsCl seed layer (**f**).

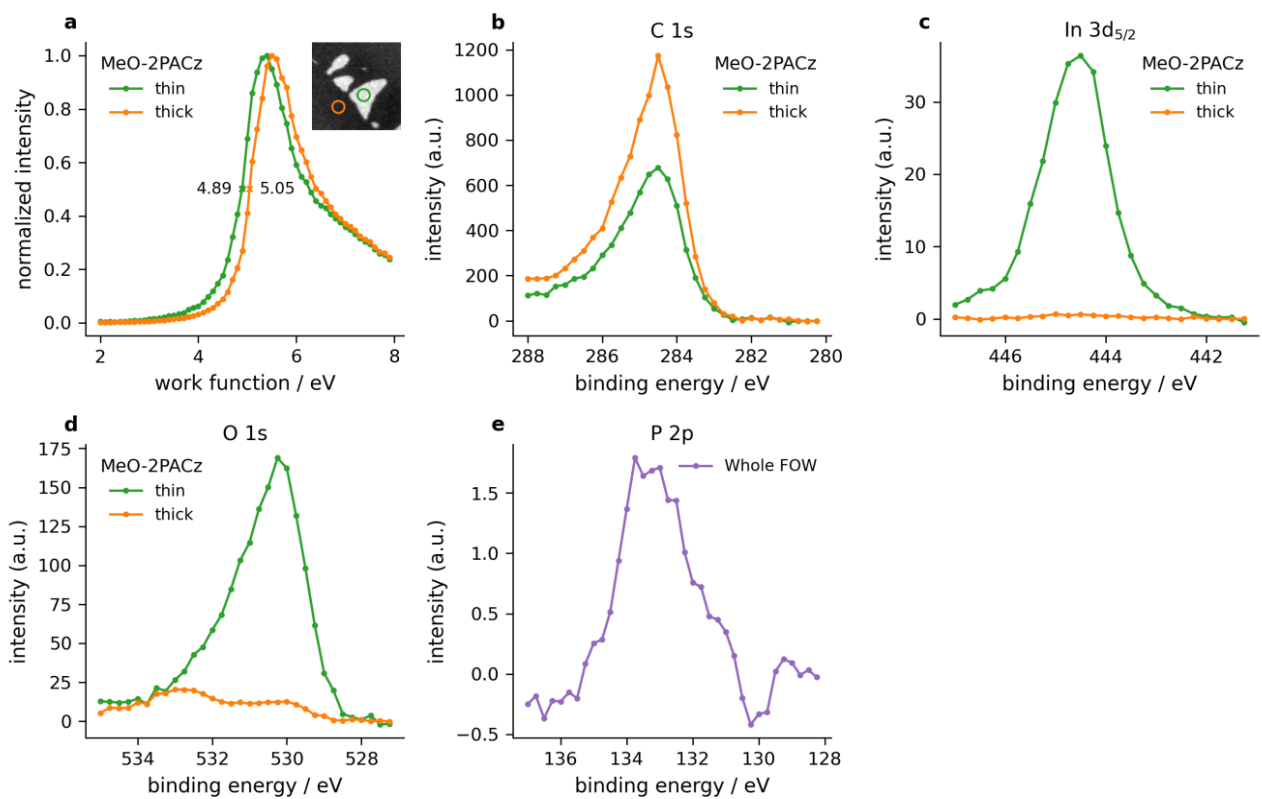

**Figure S 4**: Si/IMO:H/MeO-2PACz sample with inhomogeneous MeO-2PACz coverage. **a**, Secondary electron cutoff spectrum, obtained by averaging a kinetic energy scan in areas highlighted on a LEEM image in the inset. **b-e**, XPS spectra of C 1s (**b**), In $3d_{5/2}$ (**c**), O 1s (**d**) and P 2p (**e**) core levels on areas with thin and thick MeO-2PACz layer or the whole field of view (FOW)

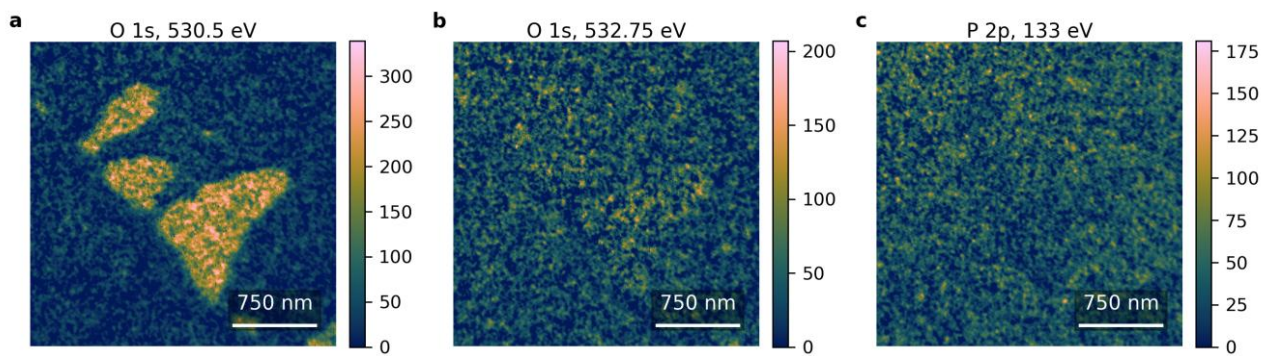

**Figure S 5**: Si/IMO:H/MeO-2PACz sample with inhomogeneous MeO-2PACz coverage. XPEEM images of O 1s, BE = 530.5 eV (**a**), O 1s, BE = 532.75 eV (**b**) and P 2p (**c**) core levels. Color scale shows XPEEM signal intensity.

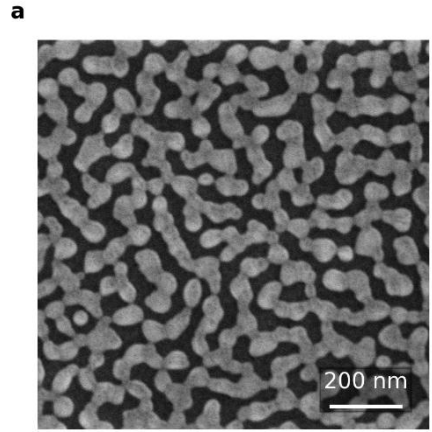

**Figure S 6**: SEM image of 10 nm CsCl layer on Si/IMO:H/MeO-2PACz sample, indicating formation of interconnected islands that do not fully cover the substrate.

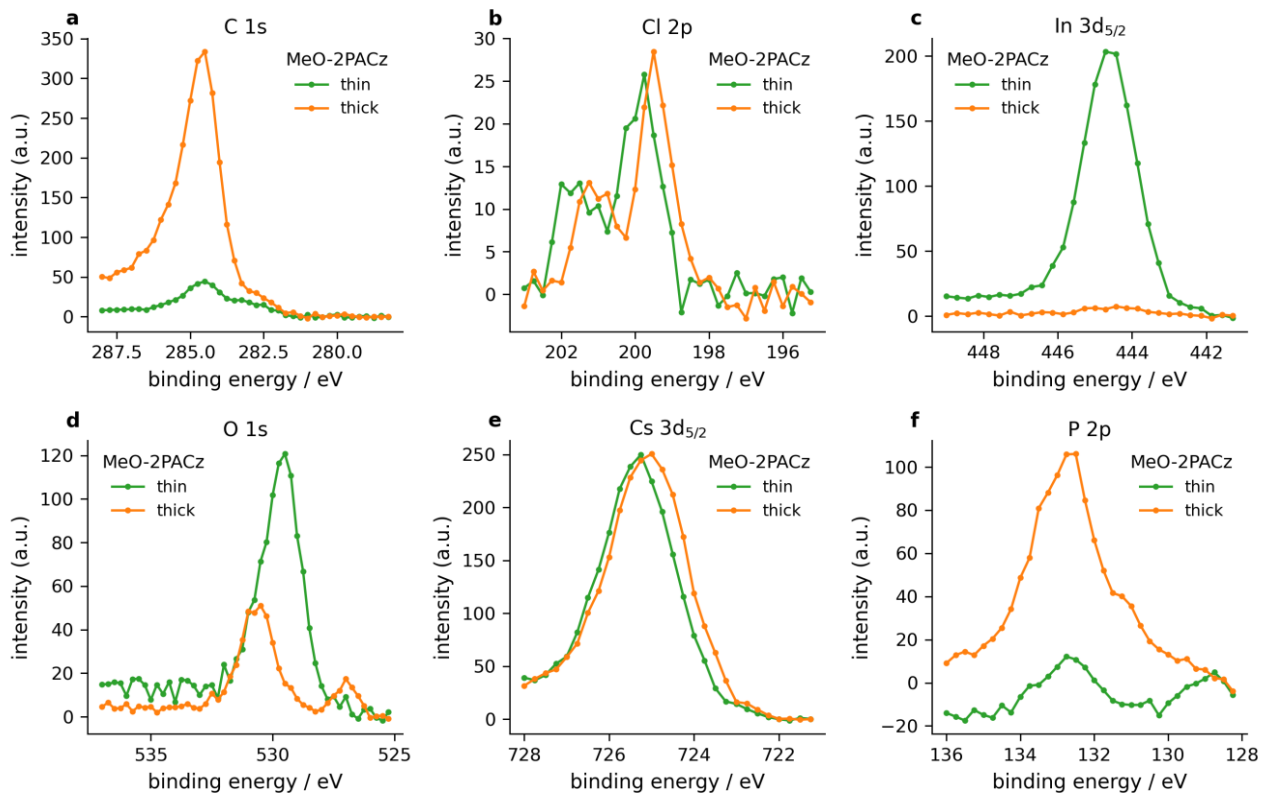

**Figure S 7**: 10 nm CsCl seed layer on Si/IMO:H/MeO-2PACz sample. XPS spectra of C 1s (**a**), Cl 2p (**b**), In $3d_{5/2}$ (**c**), O 1s (**d**), Cs $3d_{5/2}$ (**e**) and P 2p (**f**) core levels on areas with thin and thick MeO-2PACz layer.

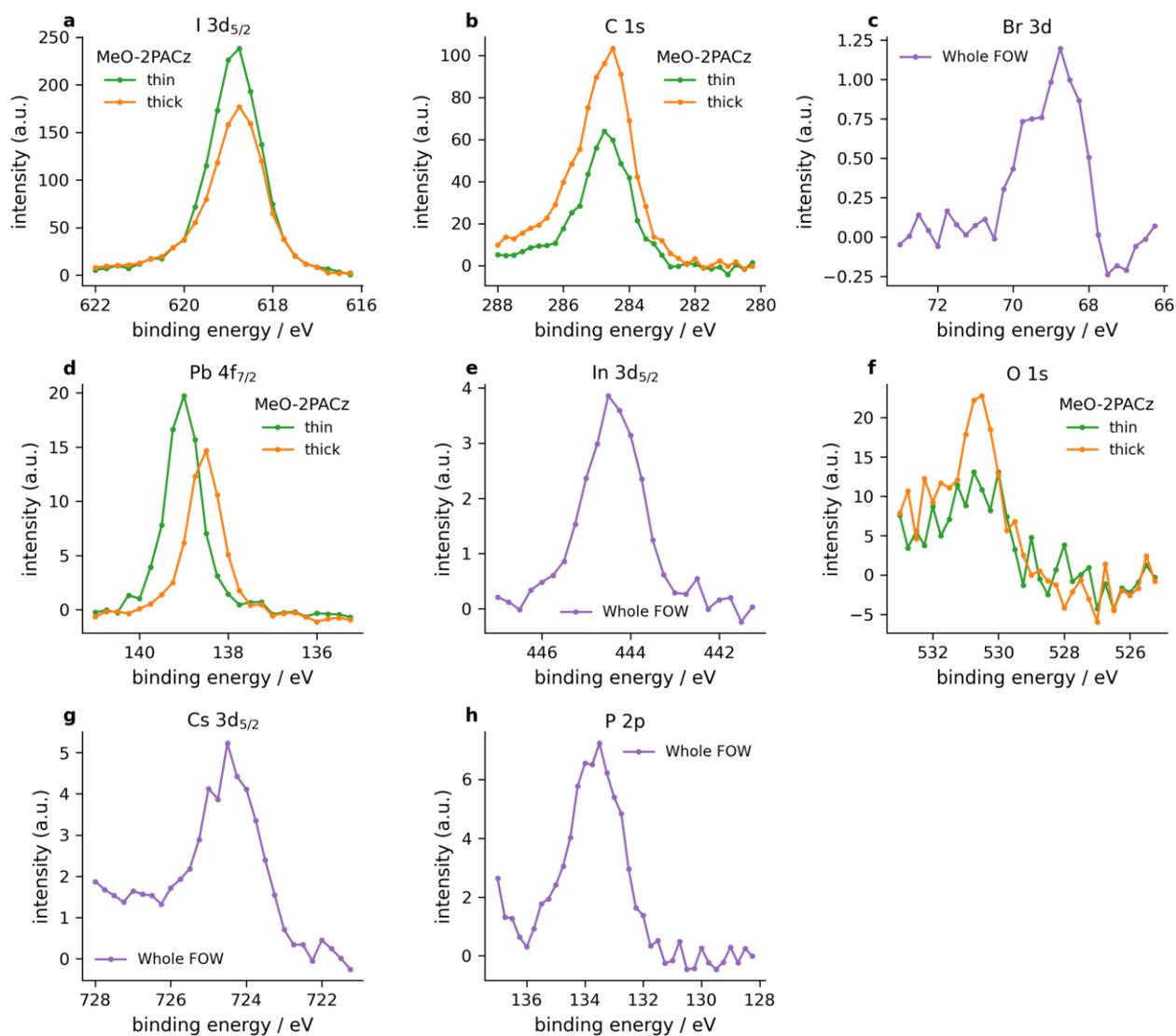

**Figure S 8**: 5 nm thick MHP on Si/IMO:H/MeO-2PACz substrate. XPS spectra of I $3d_{5/2}$ (**a**), C 1s (**b**), Br 3d (**c**), Pb $4f_{7/2}$ (**d**), In $3d_{5/2}$ (**e**), O 1s (**f**), Cs $3d_{5/2}$ (**g**), and P 2p (**h**) core levels on areas with thin and thick MeO-2PACz layer or the whole field of view (FOW).

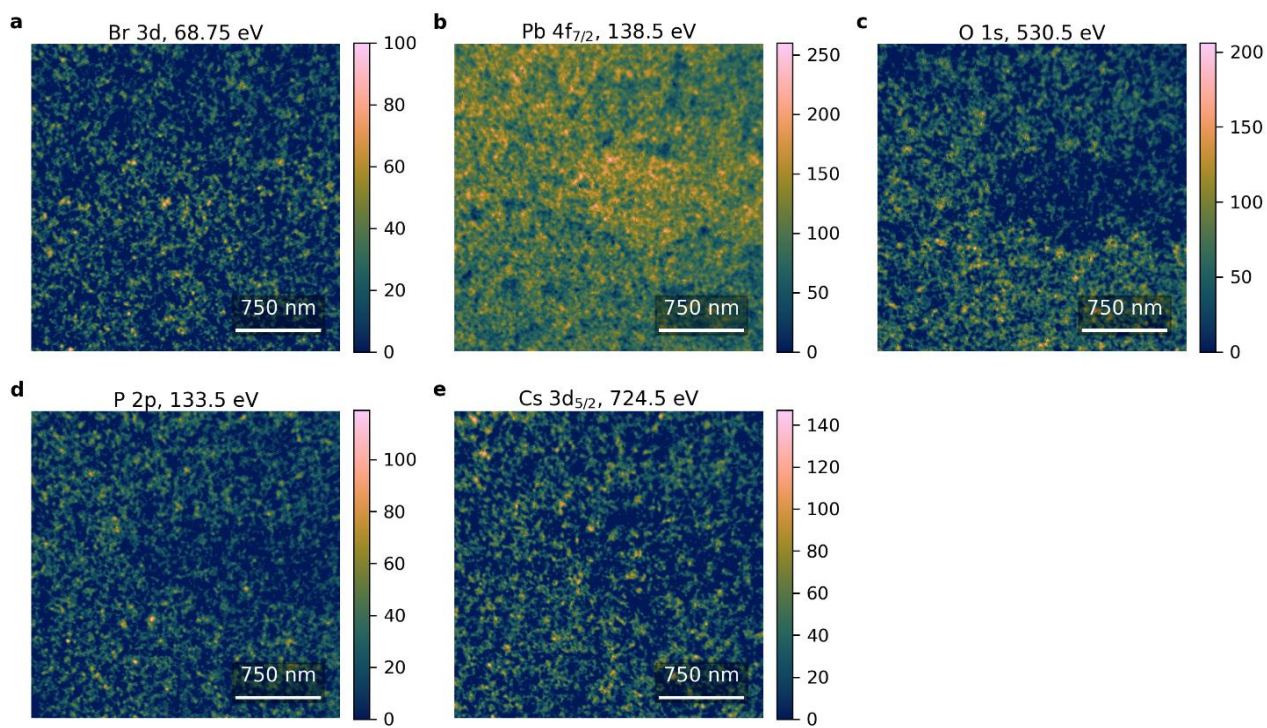

**Figure S 9**: 5 nm thick MHP on Si/IMO:H/MeO-2PACz substrate. XPEEM images of Br 3d (**a**), Pb 4f$_{7/2}$ (**b**) O 1s (**c**), P 2p (**d**), and Cs 3d$_{5/2}$ (**e**) core levels. Color scale shows XPEEM signal intensity.

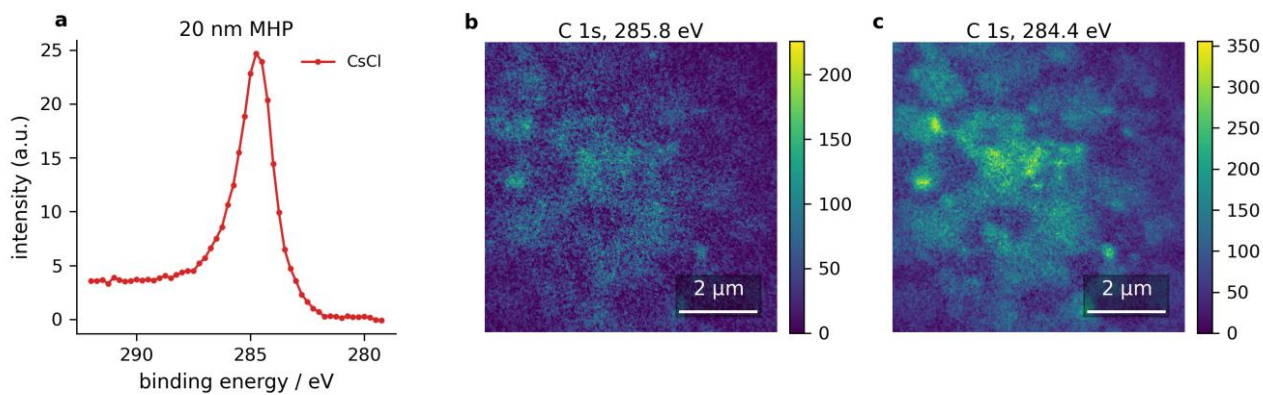

**Figure S 10**: 20 nm thick MHP on Si/IMO:H/MeO-2PACz/CsCl substrate. **a**, XPS spectra of C 1s core level. **b,c**, XPEEM images of C 1s at BE = 285.5 eV (**b**) and BE = 284.4 eV (**c**).

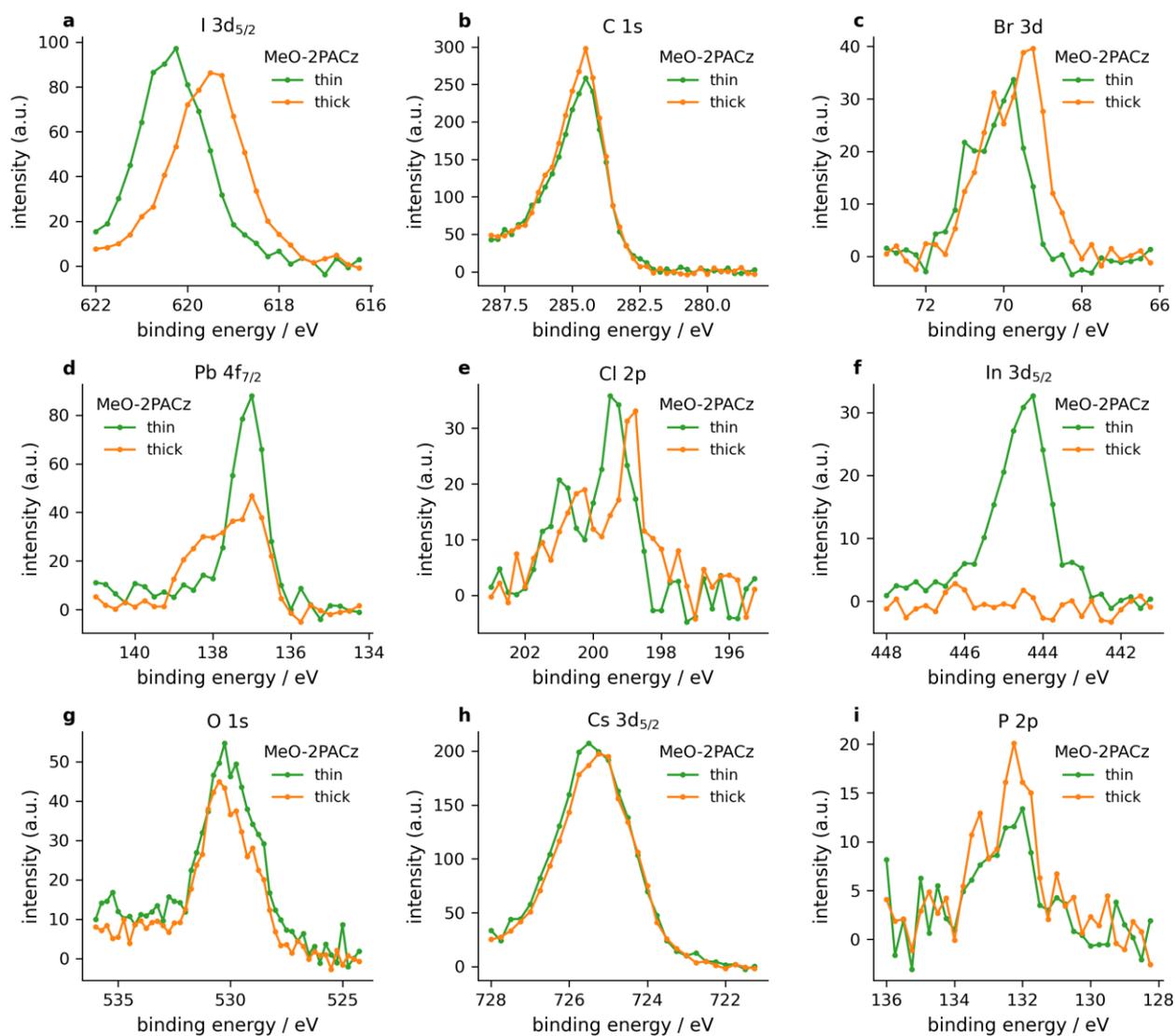

**Figure S 11**: 5 nm thick MHP on Si/IMO:H/MeO-2PACz/CsCl substrate. XPS spectra of I 3d$_{5/2}$ (**a**), C 1s (**b**), Br 3d (**c**), Pb 4f$_{7/2}$ (**d**), Cl 2p (**e**), In 3d$_{5/2}$ (**f**), O 1s (**g**), Cs 3d$_{5/2}$ (**h**), and P 2p (**i**) core levels on areas with thin and thick MeO-2PACz layer.

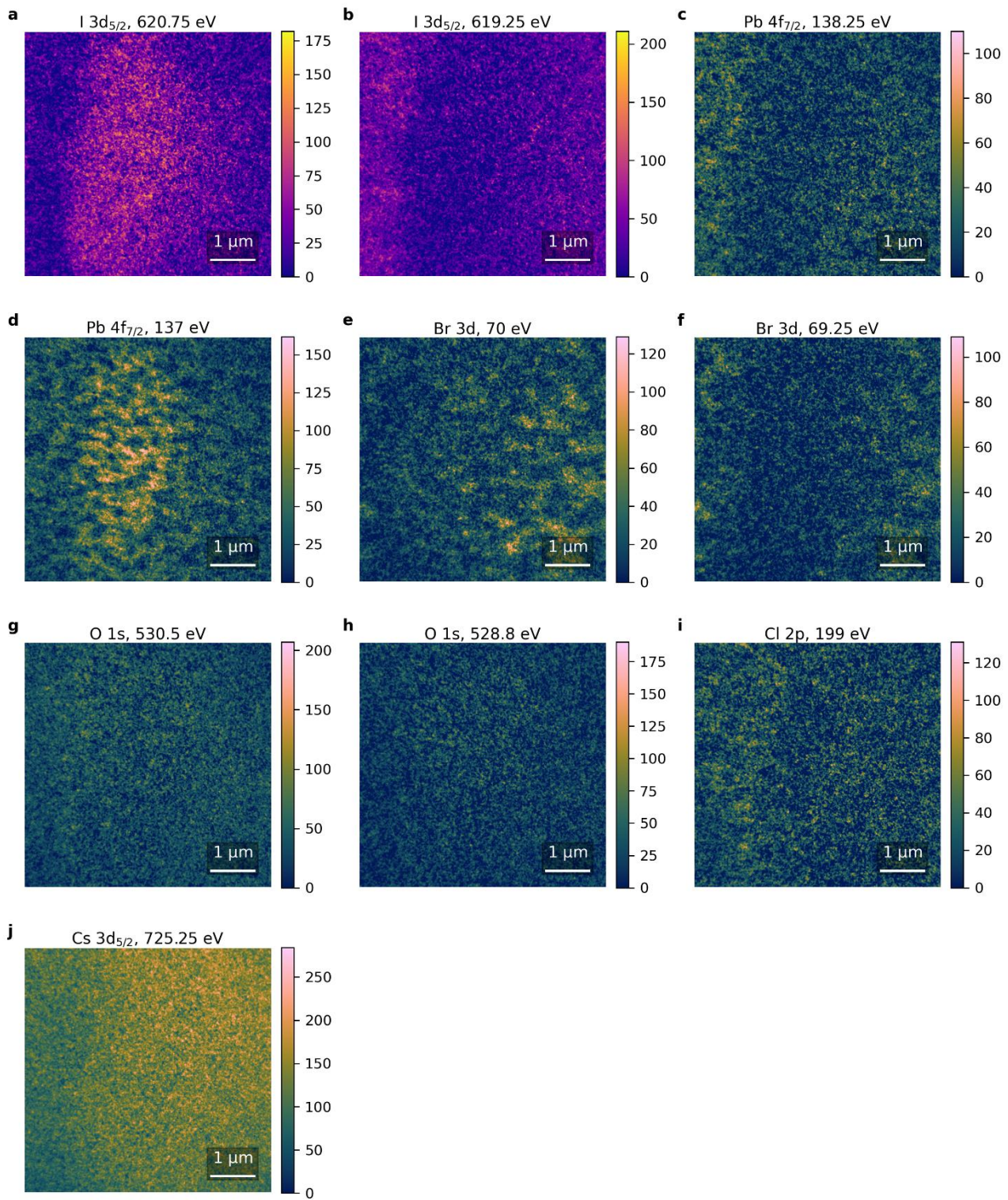

**Figure S 12**: 5 nm thick MHP on Si/IMO:H/MeO-2PACz/CsCl substrate. XPEEM images of I 3d$_{5/2}$, BE = 620.75 eV (**a**), I 3d$_{5/2}$, BE = 619.25 eV (**b**), Pb 4f$_{7/2}$, BE = 138.25 eV (**c**), Pb 4f$_{7/2}$, BE = 137 eV (**d**), Br 3d BE = 70 eV (**e**), Br 3d BE = 69.25 eV (**f**), O 1s, BE = 530.5 eV (**g**), O 1s, BE = 528.8 eV (**h**), Cl 2p (**i**), and Cs 3d$_{5/2}$ (**j**) core levels. Color scale shows XPEEM signal intensity.

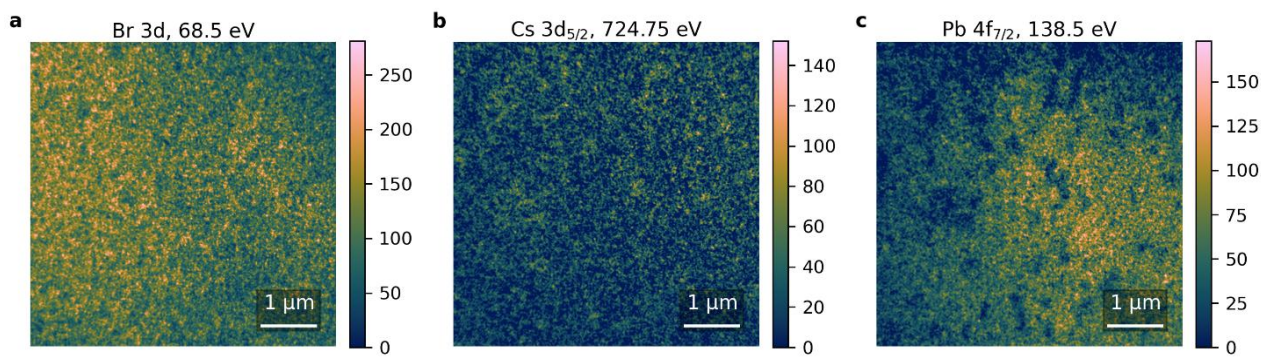

**Figure S 13**: 20 nm thick MHP on Si/IMO:H/MeO-2PACz substrate. XPEEM images of Br 3d (**a**), Cs $3d_{5/2}$ (**j**) and Pb $4f_{7/2}$ (**m**) core levels. Color scale shows XPEEM signal intensity.

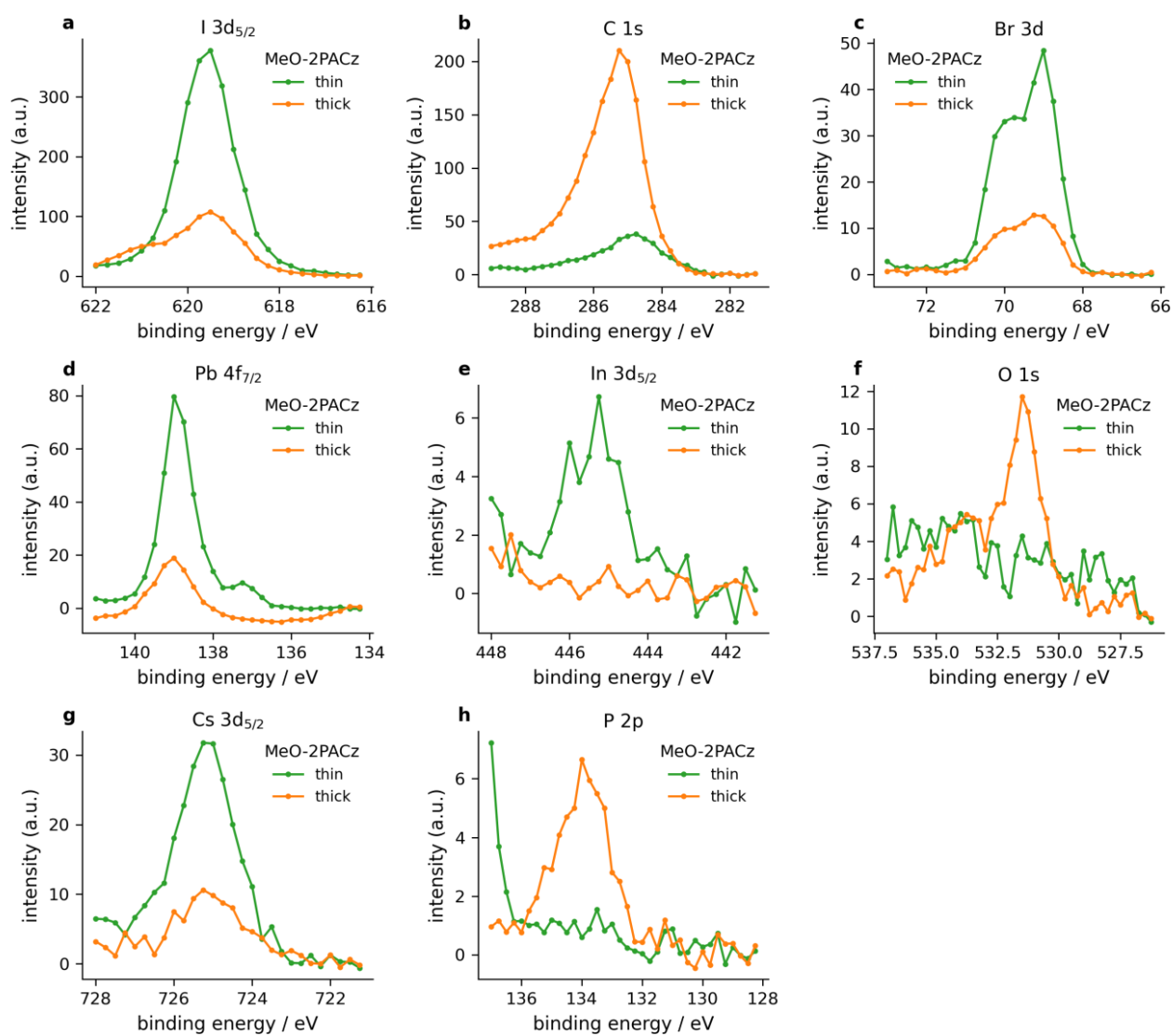

**Figure S 14**: Buried interface of 100 nm thick MHP film grown on Si/IMO:H/MeO-2PACz substrate. XPS spectra of I $3d_{5/2}$ (**a**), C 1s (**b**), Br 3d (**c**), Pb $4f_{7/2}$ (**d**), In $3d_{5/2}$ (**e**), O 1s (**f**), Cs $3d_{5/2}$ (**g**), and P 2p (**h**) core levels on areas with thin and thick MeO-2PACz layer.

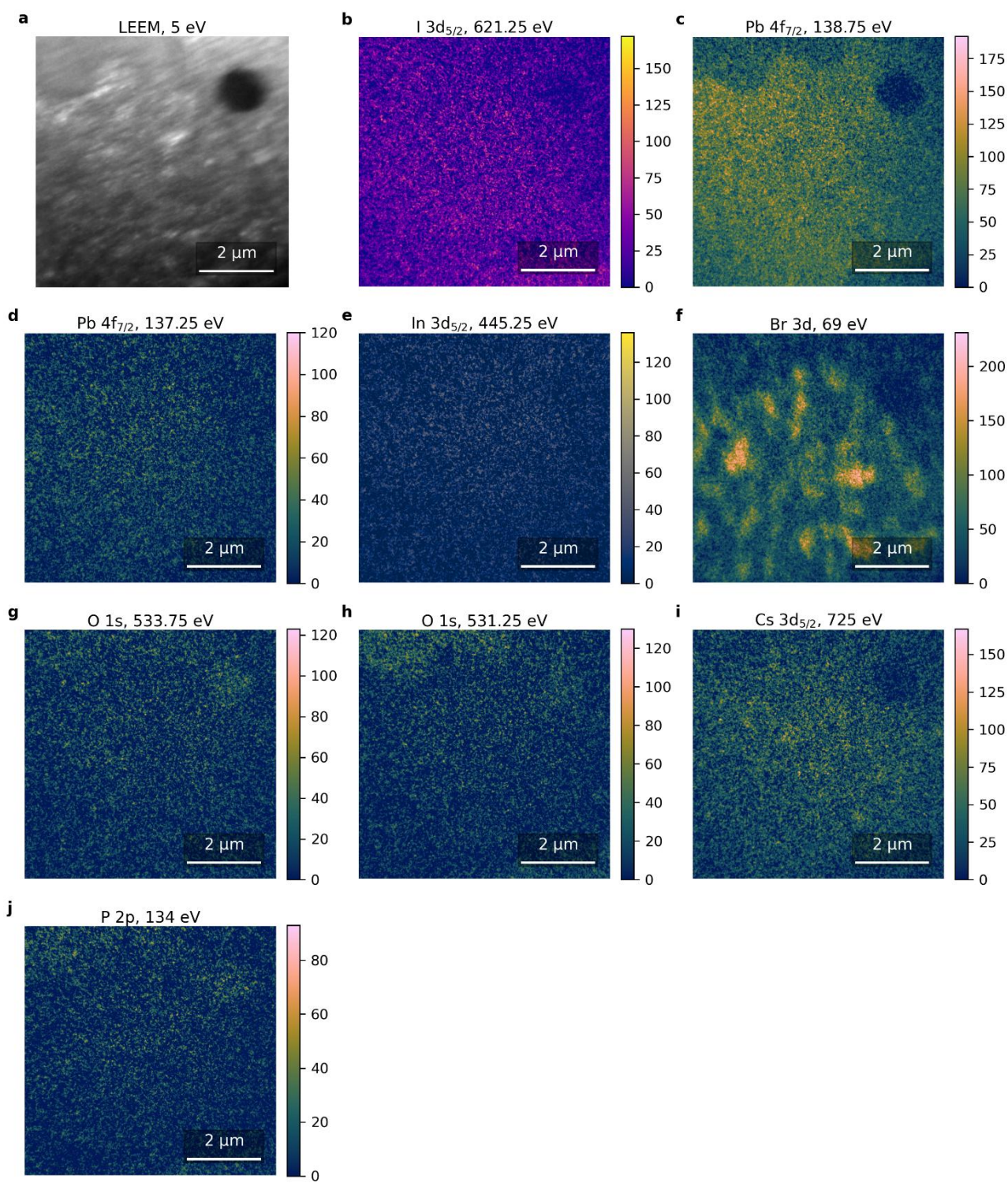

**Figure S 15**: Buried interface of 100 nm thick MHP film grown on Si/IMO:H/MeO-2PACz substrate. **a**, LEEM image. **b-j**, XPEEM images of I $3d_{5/2}$, BE = 621.25 eV (**b**), Pb $4f_{7/2}$, BE = 138.75 eV (**c**), Pb $4f_{7/2}$, BE = 137.25 eV (**d**), In $3d_{5/2}$ (**e**), Br 3d (**f**), O 1s, BE = 533.75 eV (**g**), O 1s, BE = 531.25 eV (**h**), Cs $3d_{5/2}$ (**i**), and P 2p (**j**) core levels. Color scale shows XPEEM signal intensity.

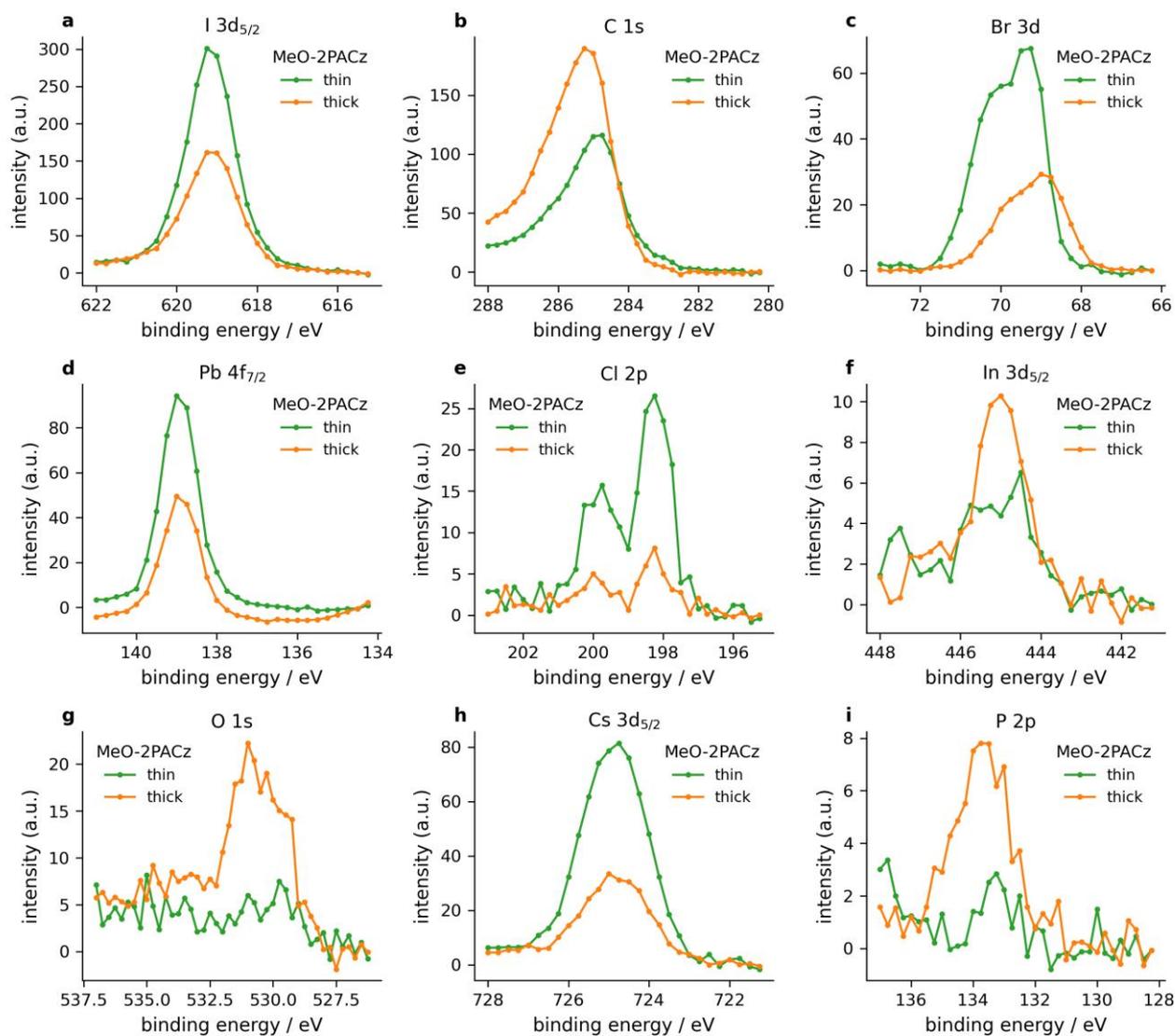

**Figure S 16**: Buried interface of 100 nm thick MHP film grown on Si/IMO:H/MeO-2PACz/CsCl substrate. XPS spectra of I $3d_{5/2}$ (**a**), C 1s (**b**), Br 3d (**c**), Pb $4f_{7/2}$ (**d**), Cl 2p (**e**), In $3d_{5/2}$ (**f**), O 1s (**g**), Cs $3d_{5/2}$ (**h**), and P 2p (**i**) core levels on areas with thin and thick MeO-2PACz layer.

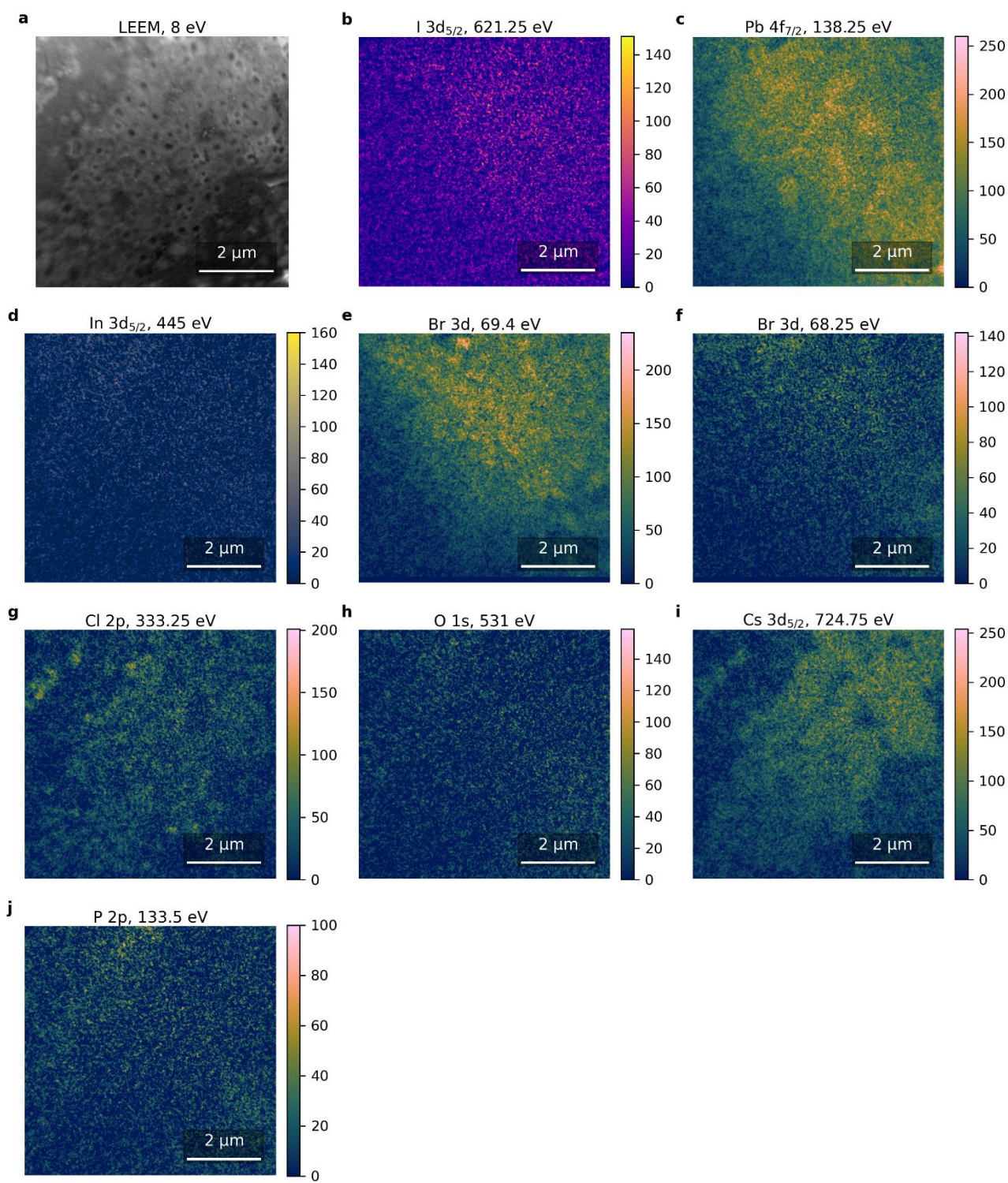

**Figure S 17**: Buried interface of 100 nm thick MHP film grown on Si/IMO:H/MeO-2PACz/CsCl substrate. **a**, LEEM image. **b-i**, XPEEM images of I 3d$_{5/2}$ (**l**), Pb 4f$_{7/2}$, BE = 138.25 eV (**b**), In 3d$_{5/2}$ (**c**), Br 3d BE = 69.4 eV (**d**), Br 3d BE = 68.25 eV (**e**), Cl 2p (**f**), O 1s (**g**), Cs 3d$_{5/2}$ (**h**), and P 2p (**i**) core levels. Color scale shows XPEEM signal intensity.

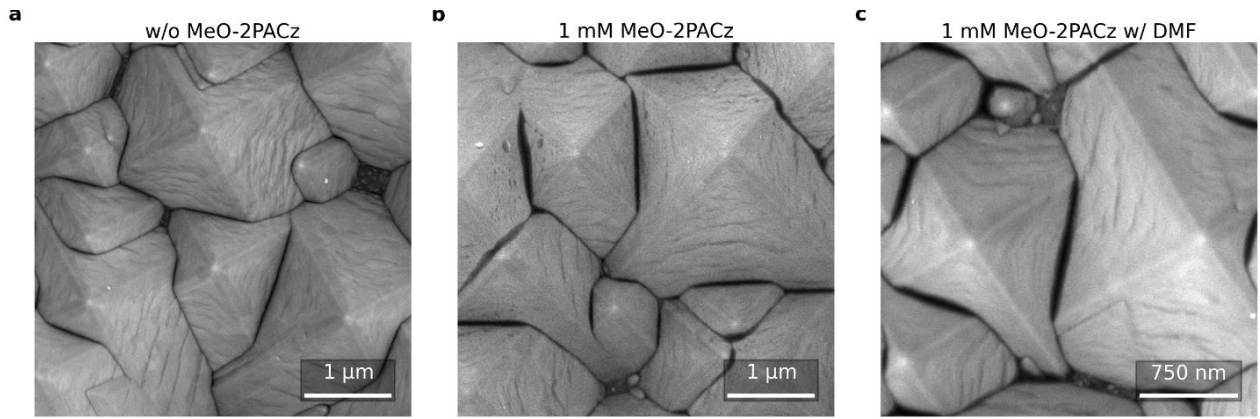

**Figure S 18**: Accumulation of spin-coated MeO-2PACz in valleys of textured silicon bottom cell. SEM images of **a**, Textured silicon bottom cell prior to deposition of MeO-2PACz, **b**, after deposition of 1 mM MeO-2PACz and **c**, after deposition of 1 mM MeO-2PACz with added DMF co-solvent.

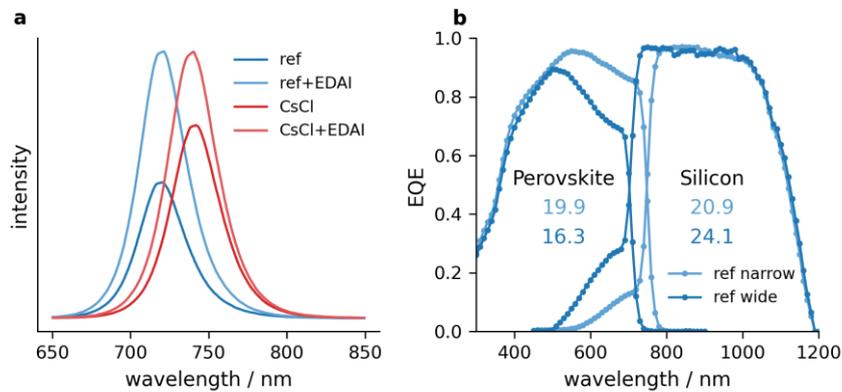

**Figure S 19**: **a**, Influence of EDAI treatment on PL of textured Si bottom cell/IMO:H/MeO-2PACz[/CsCl]/MHP substrates. **b**, Comparison of EQE spectra of narrow- and wide- band gap reference PSTs.

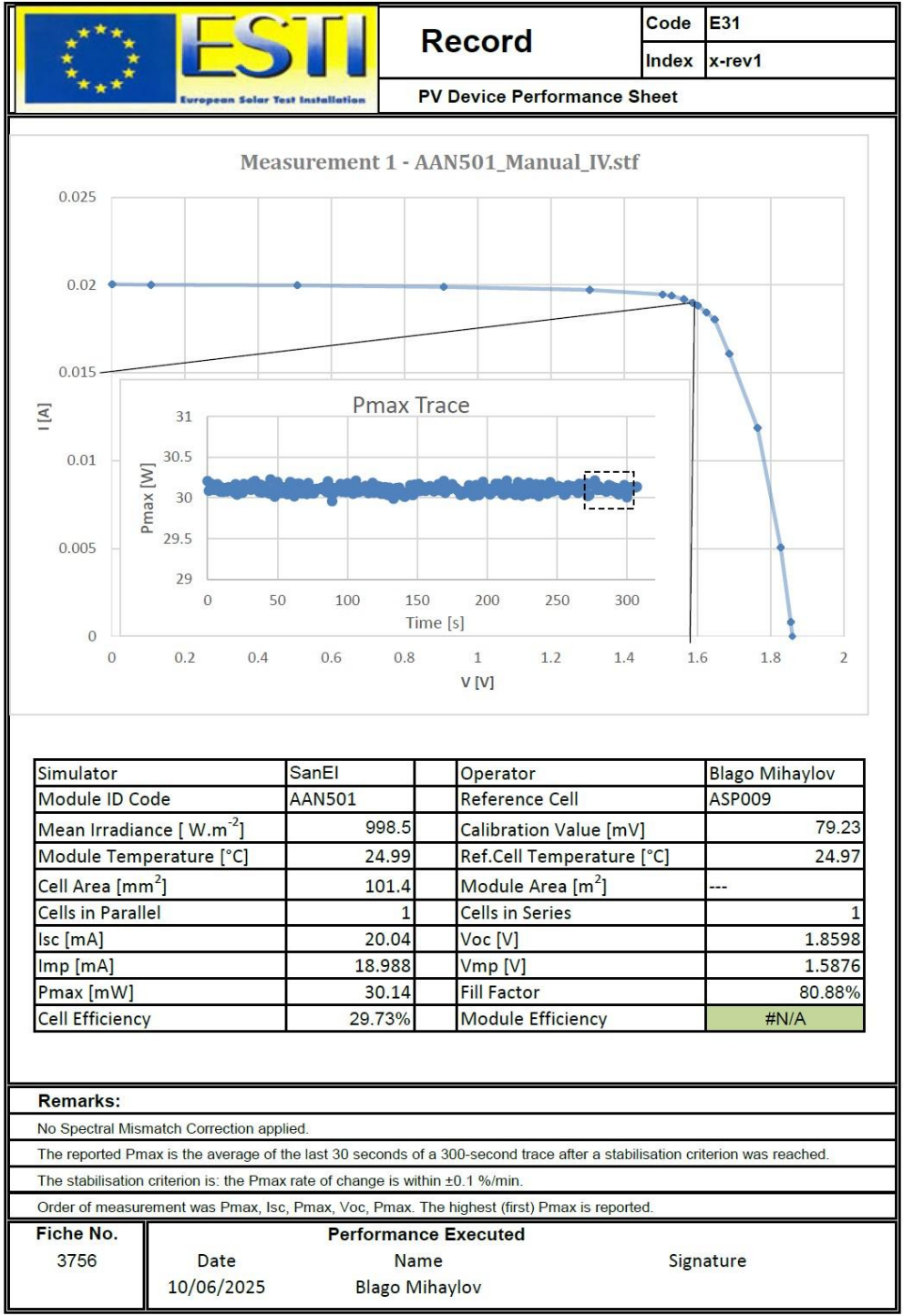

**Figure S 20**: Photovoltaic device performance sheet as certified by European Solar Test Installation (ESTI).

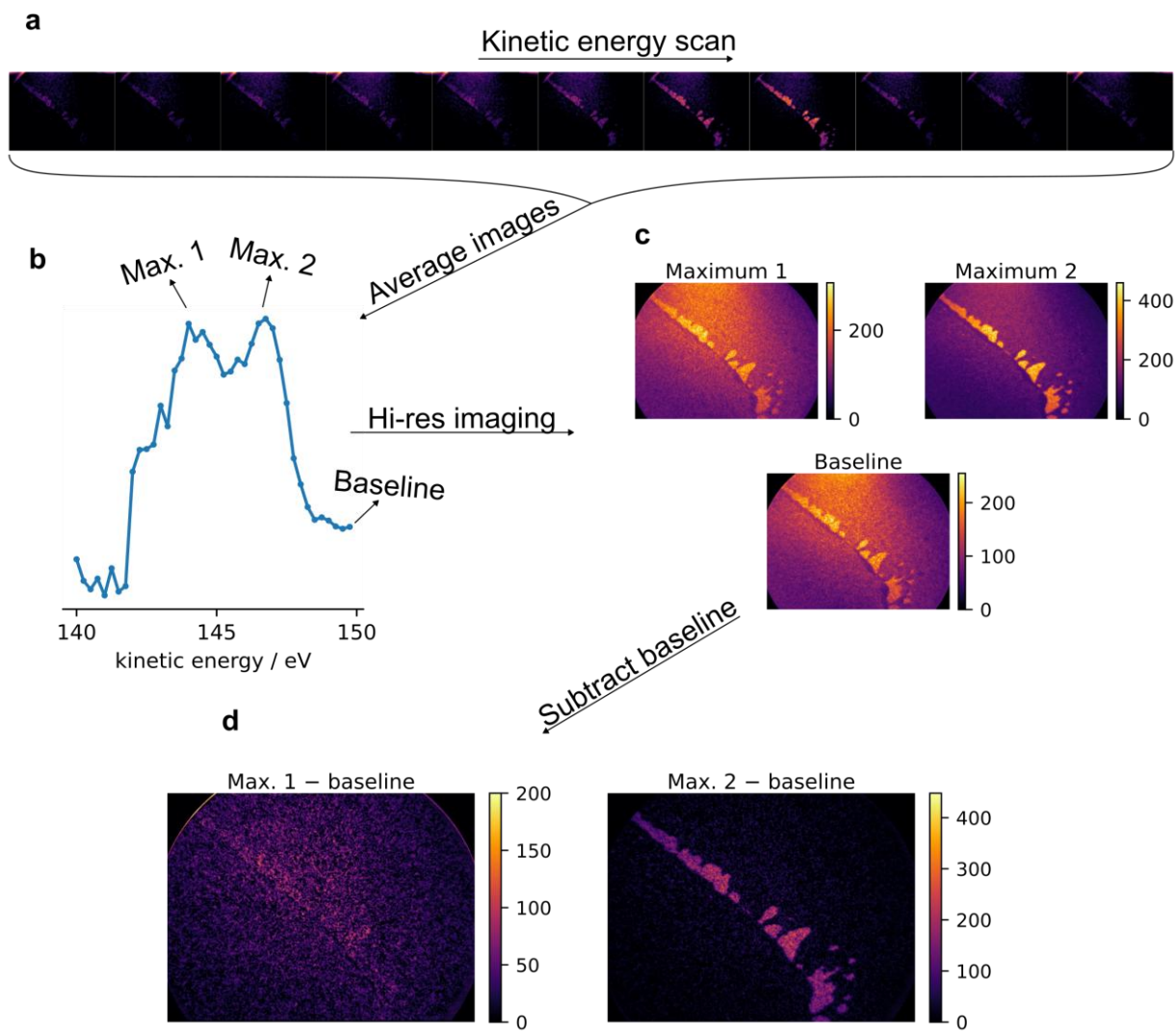

**Figure S 21**: XPEEM data processing. This example shows O 1s core level measurement (excitation energy = 680 eV). **a**, Images recorded at varying kinetic energy. Every fourth recorded image is shown here. **b**. XPS spectrum obtained as spatial average of kinetic energy scan images. **c**, High resolution images obtained at the peak maxima and baseline energy. **d**, Baseline-corrected images.